\documentclass[aps, reprint, 
superscriptaddress,
nofootinbib, notitlepage, floatfix, longbibliography]{revtex4-2}
\usepackage{gpkmac}
\usepackage{xcolor}
\usepackage{wasysym}
\usepackage{mathrsfs}
\usepackage{manfnt}
\usepackage{cleveref}
\usepackage[normalem]{ulem}

\usepackage{bbold}

\begin{document}



\title{Phase diagram of the three-dimensional subsystem toric code}

\author{Yaodong Li}
\affiliation{Department of Physics, Stanford University, Stanford, CA 94305, USA}
\affiliation{Department of Physics, University of California, Santa Barbara, CA 93106, USA}

\author{C. W. von Keyserlingk}
\affiliation{King’s College London, Strand, London WC2R 2LS, UK}

\author{Guanyu Zhu}
\affiliation{IBM T.J. Watson Research Center, Yorktown Heights, NY 10598,  USA}

\author{Tomas Jochym-O'Connor}
\affiliation{IBM T.J. Watson Research Center, Yorktown Heights, NY 10598,  USA}

\date{August 24, 2024}

\begin{abstract}

Subsystem quantum error-correcting codes  typically involve measuring a sequence of non-commuting parity check operators. They can sometimes exhibit greater fault-tolerance than conventional \emph{subspace} codes, which use commuting checks.
However, unlike subspace codes, it is unclear if subsystem codes -- in particular their advantages -- can be understood in terms of ground state properties of a physical Hamiltonian.
In this paper, we address this question for the three-dimensional subsystem toric code (3D STC), as recently constructed by Kubica and Vasmer~[\href{https://www.nature.com/articles/s41467-022-33923-4}{Nat.~Comm.~13,~6272~(2022)}], which exhibits single-shot error correction.
Motivated by a conjectured relation between single-shot properties and thermal stability, we study the zero and finite temperature phases of an associated non-commuting Hamiltonian.
By mapping the Hamiltonian model to a pair of 3D $\mb{Z}_2$ gauge theories coupled by a kinetic constraint, we find various phases at zero temperature, all separated by first-order transitions: there are 3D toric code-like phases with deconfined point-like excitations in the bulk, and there are phases with a confined bulk supporting a 2D toric code on the surface when appropriate boundary conditions are chosen. The latter is similar to the surface topological order present in 3D STC.
However, the similarities between the single-shot correction in 3D STC and the confined phases are only partial: they share the same sets of degrees of freedom, but they are governed by different dynamical rules. 
Instead, we argue that the process of single-shot error correction can more suitably be associated with a path (rather than a point) in the zero-temperature phase diagram, a perspective which inspires alternative measurement sequences enabling single-shot error correction. Moreover, since none of the above-mentioned phases survives at nonzero temperature, the single-shot error correction property of the code does not imply thermal stability of the associated Hamiltonian phase.

\end{abstract}

\maketitle


{\hypersetup{linktocpage} \tableofcontents}

\section{Introduction}

Scalable quantum computing is likely to require quantum error correction (QEC). The quest for the latter has inspired new physics over the past two decades. 
The best-known examples are topological stabilizer codes~\cite{kitaev1997}, whose associated Hamiltonian model has topological order at zero temperature.
The relation between the code and the Hamiltonian is immediate: the logical space of the code coincides with the degenerate ground space of the associated Hamiltonian.

Besides the agreement in logical space, the code model and the Hamiltonian model are physically distinct. 
For active error correction with the stabilizer code, the stabilizers are measured repeatedly in time, and the errors thus identified are removed by a (possibly nonlocal) classical decoder to push the state closer to the logical space~\cite{DKLP2001topologicalQmemory}. 
For the Hamiltonian model, the topological order is obtained only if the system is cooled to and maintained at near-zero temperature.
There are no obvious connections to be drawn between the equilibrium dynamics of the Hamiltonian (e.g. its spectrum, correlations and responses, etc) and the nonequilibrium dynamics of the code under a realistic error model and the action of a decoder.
For example, the 2D toric code can survive finite noise under repeated active correction, but the 2D toric code topological order cannot survive thermal noise at finite temperature (i.e. not a thermally stable memory)~\cite{CastelnovoChamon2D2007, nussinov-ortiz-2008-toric-code-thermal, hastings2011nonzero}.

The situation is far less clear in the case of \emph{subsystem} codes~\cite{poulin2005subsystemcode, bacon2006compass}, where the check operators to be measured do not necessarily mutually commute, but can nevertheless be used to preserve logical information.
The associated Hamiltonians are thus not sums of commuting projectors, and can potentially host a nontrivial phase diagram.
A basic question is whether it is possible to still attribute certain properties of the code to the Hamiltonian in equilibrium, and if possible, to which phase.

Subsystem codes are of great practical interests.
The three-dimensional gauge color code (GCC)~\cite{bombin2015gaugecolorcode, bombin2016dimensionaljump, bombin2015singleshot} is an example that stands out, for two reasons, namely universal fault-tolerant gates~\cite{bombin2015gaugecolorcode, bombin2016dimensionaljump, kubica2015universalgatecolorcode}, and a property called ``single-shot error correction''~\cite{bombin2015singleshot, BrownNickersonBrowne2016GCC}; see also~\cite{fawzi2018quantumexpandercodes,campbell2019theorysingleshot,breuckmann2021single, quintavalle2021singleshot, higgott2022improved}.
A single-shot quantum memory is one for which syndrome error correction operations can be performed at a constant rate independent of the code distance, despite the possibly noisy measurements.
For the GCC, its single-shot error correction is achieved by building redundancy into the check operators, which guarantees that the syndromes form extended, closed loops in the absence of syndrome error.
This property is known as \textit{syndrome confinement}, or simply \textit{confinement}.
We describe these in more detail in Sec.~\ref{sec:background_ssec}.

Resilience to imperfect measurements also occurs in the 4D toric code, which is self-correcting~\cite{DKLP2001topologicalQmemory} and also has single-shot properties.
Both properties are consequences of closed, loop-like excitations/syndromes,
which cost energies linear in their lengths.
The similarities of error syndromes between the 3D GCC and the 4D toric code have led to conjectures~\cite{bombin2015gaugecolorcode, brown2016selfcorrectingRMP, preskill2017qec} that an associated \textit{quantum} Hamiltonian of the GCC may have a topologically ordered phase that is self-correcting.
This is an intriguing conjecture for (i) a self-correcting memory in 3D would be quite surprising given that previous attempts~\cite{haah2011cubic, bravyihaah2013, michnicki20123d, michnicki2014prl, brown2016selfcorrectingRMP} have not been successful, and (ii) it hints at certain general connections between single-shot error correction -- which is a property of the nonequilibrium dynamical of the code -- and thermodynamic phases of the Hamiltonian.


Recently, a model closely related to the GCC, namely the 3D subsystem toric code (3D STC), was introduced by Kubica and Vasmer~\cite{KubicaVasmer}.\footnote{The 2D STC was introduced much earlier, in Ref.~\cite{bravyi2012twodimSTC}, and was shown to be equivalent to the 2D toric code up to a finite depth local circuit.}
This model is in many ways qualitatively similar to the GCC -- for example, single-shot error correction with the STC is analogous to the GCC, and it is possible to perform universal fault-tolerant computation with the STC~\cite{IversonKubica, iverson2020aspects} -- but on a simpler lattice.
One can similarly conjecture the existence of a self-correcting stable quantum memory in an appropriate Hamiltonian for this code.


In this paper, we study a family of concrete Hamiltonians for the 3D STC.
They are designed to capture the most important properties of the code while being as simple as possible.
In particular, redundancies among the check operators (which leads syndrome confinement) are built into the Hamiltonian as kinetic constraints.
By studying its phase diagram, we find phases equivalent to the 3D toric code or the 2D toric code.
They are known to be not self-correcting, and are unstable at finite temperatures.
We therefore conclude that syndrome confinement alone is not a sufficient mechanism for thermal stability or self-correction.

We further find that single-shot error correction of the 3D STC cannot be attributed to any particular phase, and must be understood as a property of a path through the phase diagram.
Our results suggest that the relation between error correction and equilibrium  phases of the same code is a subtle matter; in particular, the ``error correcting'' phase of the code may be suitably understood as an intrinsically non-equilibrium phase that has no equilibrium analogs. 

In the remainder of this section, we provide necessary background and motivation for our work, before giving a more detailed overview of our results.



\subsection{Self-correction and thermal stability}

``Self-correcting quantum memories''~\cite{brown2016selfcorrectingRMP} refers to a class of quantum many-body Hamiltonians which can be used to store quantum information reliably when in contact with a noisy environment, without the need of active error correction.
They are therefore practically desirable.

More precisely, consider a Hamiltonian with a degenerate ground space.
After an arbitrary initial pure state within the ground space is prepared, the system is put into contact with an environment, described by local physical interactions between degrees of freedom in the system and those in the environment.
After evolution for time $t$, one is allowed to perform a recovery operation, typically in the form of measuring local, few-body operators and a feedback unitary as determined by the measurement results.
The recovery is successful if its result coincide with the initial state.
For a finite $L$, the memory time $t_{\rm mem}(\epsilon, L)$ of the Hamiltonian is defined as the maximum $t$ above which the recovery fails with probability greater than $\epsilon$.
The Hamiltonian is said to be self-correcting if $ \lim_{L \to \infty} t_{\rm mem}(\epsilon, L) = \infty$ for any $\epsilon > 0$.

Implicit in this definition is often the requirement that the Hamiltonian is defined on a Euclidean lattice in finite dimensions, with a finite number of degrees of freedom per site.
The interactions in the Hamiltonian are also required to be short ranged.

One well known example of a self-correcting memory is the 4D toric code~\cite{DKLP2001topologicalQmemory, Alicki_2010}.
In this model, the excitations are closed loops, whose energy grows linearly with their lengths.
On the other hand, the occurrence of a logical error involves the creation of such loops of linear dimension $O(L)$.
Therefore, the Hamlitonian is said to have a macroscopic energy barrier to logical errors, as is crucial for its self-correction.


For the purpose of the current paper, it is useful to compare the 4D toric code with the 2D toric code.
The latter has point-like excitations which are ``deconfined'', meaning that the excitation energy of a pair of excitations is a constant, independent of  their separation.
Consequently, the 2D toric code only has an $O(1)$ energy barrier, and have been shown to have a finite memory time~\cite{2009JPhA...42f5303A}, therefore not a self-correcting memory.
The 3D toric code also hosts deconfined point-like excitations and is also believed to have a finite memory time, for similar reasons.

Attempts have been made in searching for self-correcting quantum memories below four dimensions.
In 2D, no-go results have been obtained~\cite{bravyi2008nogo, yoshida2011feasibility, landon2013local}.
However, in 3D, interesting examples include Haah's cubic code~\cite{haah2011cubic} and Michnicki's welded code~\cite{michnicki20123d}, which have $O(\log L)$ and $O(L^\alpha)$ energy barriers to logical errors, respectively.
Despite this, they have subsequently be shown to be not self-correcting~\cite{bravyihaah2013, 2017PhRvA..95c2324S}.
The existence of 3D self-correcting memories remains an open question.

Closely related to the notion of self-correction is \textit{thermally stable topological order}, as defined for the Gibbs states of the Hamiltonian.
For toric codes in 2D and 3D, no topological order at finite temperature was found~\cite{CastelnovoChamon2D2007, CastelnovoChamon3d2008, nussinov-ortiz-2008-toric-code-thermal, 
iblisdir_thermal_2010,
hastings2011nonzero, yoshida2011feasibility, 2009JPhA...42f5303A, 2017PhRvA..95c2324S}, whereas similar analysis for the 4D toric code~\cite{peterlu2020negativity, zackweinstein2019} find a nonzero critical temperature.
The two notions seem to agree for these examples, but no general relations between the two are known.
We refer the reader to Ref.~\cite{brown2016selfcorrectingRMP} for a more detailed discussion.

\subsection{Single-shot error correction with the gauge color code and Bomb{\'\i}n's conjecture \label{sec:background_ssec}}

Single-shot error correction is a property of quantum memories that allows error correction after \textit{every} $O(1)$ rounds of noisy syndrome measurements.
It is a property desirable for logical gates~\cite{bombin2015gaugecolorcode}, as the gate time is often limited by the rate at which error correction can be carried out~\cite{terhal2015RMP}.
Recall that for the 2D toric code, $O(L)$ rounds are necessary before decoding for the threshold to be nonzero~\cite{DKLP2001topologicalQmemory}; performing error correction too frequently based on faulty measurements tends to introduce more physical errors.

We represent the process of single-shot error correction as follows, following the approaches taken in Refs.~\cite{bombin2015singleshot, KubicaVasmer}.
Let $\Pi$ be the projection operator onto the code space, $\mathbb{N}_{\tau}$ be the set of physical error channels parameterized by a physical error rate $\tau$, and $\mathbb{R}_\eta$ be the set of recovery channels based on syndrome measurements with error rate $\eta$.
We can write
\begin{align}
    \mathbb{R}_0 \circ (\mathbb{R}_\eta \circ \mathbb{N}_{\tau})^t  \circ \Pi \subseteq \mathbb{N}^{\mathcal{L}}_{\epsilon(t)} \circ \Pi.
\end{align}
That is, after $t$ rounds of physical errors in conjunction with noisy recovery, and a final round of perfect measurements and recovery, the state is returned to the code space.
Its effect can therefore be captured by a logical error channel acting within the code space (the set of such channels is denoted $\mathbb{N}^{\mathcal{L}}_{\epsilon(t)}$) where $\epsilon(t)$ parameterizes the logical error rate.
For a finite size system, we can define $t_{\rm mem}(\epsilon, L)$ similarly as before.
The code is a single-shot memory if there exists $\eta_{\rm th} > 0$, $\tau_{\rm th} > 0$ such that for all $\eta < \eta_{\rm th}$ and $\tau < \tau_{\rm th}$, $\lim_{L\to\infty} t_{\rm mem}(\epsilon, L) = \infty$ for any $\epsilon > 0$.

The 4D toric code is believed to be a single-shot memory~\cite{DKLP2001topologicalQmemory}.
Due to the excitations being closed contiguous loops, a decoder can take advantage of this geometric constraint and act locally in contracting the loops, despite measurement errors which appear as missing segments of the loops~\cite{kubicapreskill2019, Vasmer_2021}.
No positive results on single-shot error correction for 2D and 3D toric codes are known, as point-like syndromes are local (rather than extended), and no obvious such constraints can be exploited.



Remarkably, there are single-shot memories in 3D.
The first such example is the gauge color code by Bomb{\'\i}n~\cite{bombin2015gaugecolorcode, bombin2015singleshot}, and more recently Kubica and Vasmer constructed the the 3D subsystem toric code~\cite{KubicaVasmer}, which can be viewed as a simplification of the gauge color code.
Both are \textit{subsystem} codes, which utilize noncommuting check operators.
By relaxing the requirement of commuting checks, more redundancy among them can be built into the construction, thereby leading to greater resilience against measurement errors.

We will discuss the 3D subsystem toric code in much more detail in the following sections.
Here, it suffices to note the most important property shared by both codes, namely \textit{syndrome confinement}: due to nontrivial redundancies between the checks, the syndromes for $X$ and $Z$ errors are both extended closed loops, in contrast to the 3D toric code, where one type of error leads to point like syndromes.
Faulty measurements appear as missing segments of such flux loops.
The segments are typically short, and can be correctly distinguished from true stabilizer errors by the decoder.
This leads to a reliable readout of the syndrome using a single round of imperfect measurements, without the need for repetition~\cite{bombin2015gaugecolorcode, bombin2015singleshot, BrownNickersonBrowne2016GCC, KubicaVasmer}.

Based on the similarity in its syndromes with the 4D toric code, in particular syndrome confinement, Bomb{\'\i}n conjectured~\cite{bombin2015gaugecolorcode, bombin2015singleshot} that an appropriate Hamiltonian associated to the gauge color code might have a phase that is self-correcting.

Along these lines, it was pointed out previously that the gauge color code is a self-correcting memory when one-form symmetries are strictly imposed~\cite{RobertsBartlett, RobertsYoshidaKubicaBartlett, StahlNandkishore}.
However, it is unclear if the symmetries can naturally emerge from a microscopic Hamiltonian, and no results are known when these symmetry constraints are removed.

\subsection{Summary of our results}


With these motivations, we address the conjecture in the context of the 3D subsystem toric code (3D STC).
The rest of the paper is organized as follows.

We first write down a family of translationally invariant Hamiltonians (Eq.~\eqref{eq:Hamiltonian_J_K} of Sec.~\ref{sec:bulk_phases}) with non-commuting terms that aim to capture the most important features of the code, and from which we can extract meaningful conclusions despite being not exactly solvable.
Then, we explore the phase diagram of this Hamiltonian at both zero and nonzero temperatures.
As a general strategy, 
we first focus on zero temperature.
After understanding the zero temperature phase diagram, we can discuss the thermal stability of each of the phases.

It is worth noting that no one-form symmetries are imposed by hand in constructing the Hamiltonian.
Generators of the one-form symmetries are precisely the stabilizers of the subsystem code~\cite{RobertsBartlett, RobertsYoshidaKubicaBartlett, StahlNandkishore}, which have finite (rather than infinite) coefficients in our Hamiltonian.

In Sec.~\ref{sec:TC_limit}, we first focus on limiting cases where the Hamiltonian  contains only commuting terms.
In these limits, we find zero temperature phases that are equivalent to the 3D toric code, or its electromagnetic dual.

In Sec.~\ref{sec:lattice_duality}, we move away from the above limit, and address the full Hamiltonian with non-commuting terms.
An important technical observation is a lattice duality transformation which maps the STC Hamiltonian to two copies of 3D $\mb{Z}_2$ lattice gauge theories (LGT), whose phase diagram at zero temperature is well known.


With these, we find that for any parameter choice that respects translation symmetry and an exchange symmetry between the two LGTs, the system is either in the topologically ordered phase of the 3D toric code, or its electromagnetic dual (Fig.~\ref{fig:ZN_phase_diagram}).
The two phases are separated by a first-order phase transition.
Furthermore, we note that neither of the phases are stable at finite temperatures.

Crucial to the Hamiltonian is a kinetic constraint that point charges must emanate fluxes in both LGTs.
This is the property that leads to syndrome confinement, but so far it played no role in the phase diagram.


In Sec.~\ref{sec:boundary_surface_code}, we consider breaking the exchange symmetry between the two LGTs, and putting them into mutually dual phases.
The kinetic constraint leads to the confinement of all bulk excitations.
This explains the absence of ground state degeneracy (and topological order) when the underlying lattice does not have a boundary.

Thus, to properly capture the code properties, it is necessary to consider a lattice with a boundary.
In Sec.~\ref{sec:bc_microscopics},  we describe the microscopic boundary conditions from Ref.~\cite{KubicaVasmer}, where a 2D toric code is found on the boundary, in addition to the confined bulk.
In Sec.~\ref{sec:bc_fluxes}, we give a coarse-grained description of such boundary conditions, rephrasing known microscopic facts about the STC in fairly general terms of the gauge theory.
For example, it becomes clear that the ground state degeneracy of the 3D STC really comes from the boundary 2D toric code, and the logical operator of the former is bounded by that of the latter.
We can also reverse-engineer the microscopic boundary conditions from their gauge theory interpretation.
In Sec.~\ref{sec:bare_logicals}, we further construct membrane bare logical operators of the code from the gauge theory picture.
They are symmetries of the Hamiltonian, and guarantees a ground state degeneracy throughout the phase diagram, but are not previously understood.
They are also crucial to performing logical operations on the encoded qubits, in particular code switching~\cite{bombin2016dimensionaljump}.


The phase diagram of the model is shown in Fig.~\ref{fig:offdiagonal_phase_diagram}.
The phases we find are equivalent to toric code phases (either in 2D or 3D) up to a finite-depth unitary circuit; and as is well known for 2D and 3D toric codes
they are not stable at finite temperatures~\cite{CastelnovoChamon2D2007, CastelnovoChamon3d2008, nussinov-ortiz-2008-toric-code-thermal, 
iblisdir_thermal_2010,
hastings2011nonzero, yoshida2011feasibility, 2009JPhA...42f5303A, 2017PhRvA..95c2324S}.
We thus conclude that the STC Hamiltonian in Eq.~\eqref{eq:Hamiltonian_J_K} cannot serve as a thermally stable quantum memory, and that syndrome confinement alone is not a sufficient mechanism for thermal stability or self-correction.

We expect similar conclusions to hold for the 3D gauge color code, for the latter can also be described by $\mb{Z}_2$ gauge theories coupled by kinectic constraints~\cite{kubica2018ungauging}.

In Sec.~\ref{sec:ssec}, we discuss the relation between the thermodynamic phases and single-shot error correction.
Partial analogies can be made between the bulk-confined phases and single-shot error correction, but we emphasize important differences and that they should be treated as different phenomena.
Instead, we argue that the process of single-shot error correction is most suitably interpreted as a sequence of transitions in the phase diagram induced by check measurements.

Finally, in Sec.~\ref{sec:discussion} we summarize our results, and discuss a few future directions.

\section{The subsystem toric code Hamiltonian \label{sec:bulk_phases}}

\begin{figure*}[!ht]
    \includegraphics[width=.8\textwidth]{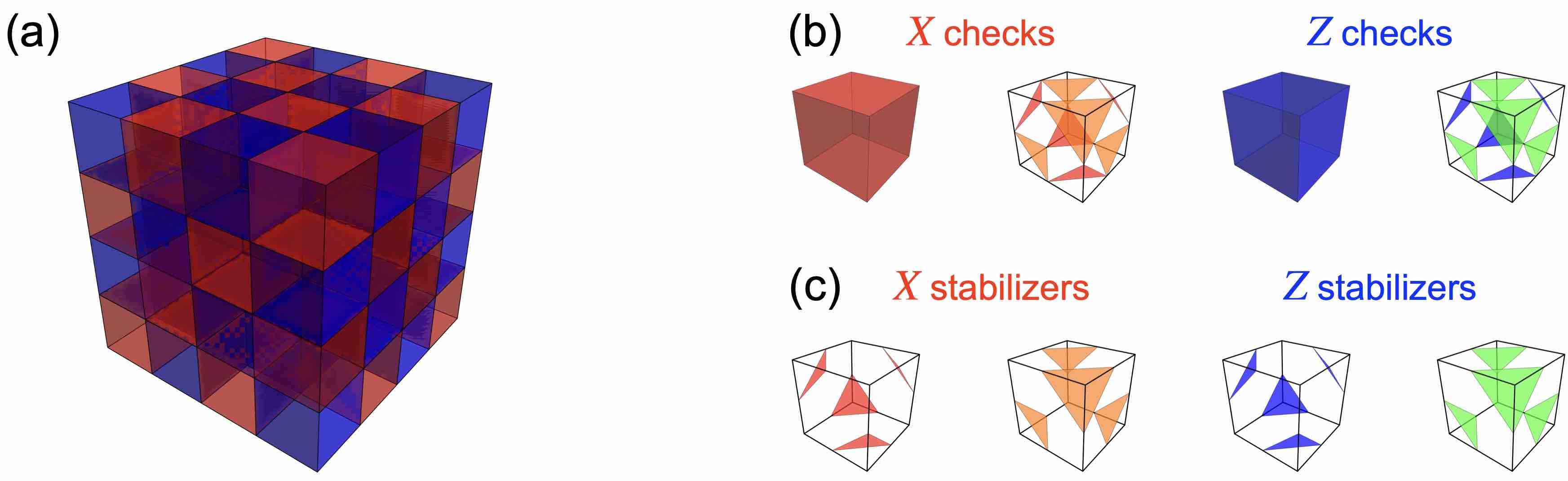}
    \caption{Checks and stabilizers of the 3D STC in the bulk~\cite{KubicaVasmer}.
    (a) The model is defined on a ``checkboard'' cubic lattice, with alternating blue and red colors on nearest-neighbor cells.
    The qubits live on the midpoints of the edges.
    (b) Within each red cell we have eight 3-qubit $X$ check operators, which can be partitioned into two groups of four non-overlapping checks, which we color red (R) and yellow (Y).
    Similarly for each blue cell, we group the checks into two colors, blue (B) and green (G).
    The colors are chosen such that the blue checks commute with yellow checks, and red checks commute with green checks.
    (c) There is one stabilizer associated to each cell, which can obtained as the product of all checks of the same color within the cell.
    }
    \label{fig:lattice}
\end{figure*}

The 3D STC is defined on a 3D cubic lattice, with qubits living on the midpoints of the edges.
The cells are colored in a checkerboard pattern, see Fig.~\ref{fig:lattice}(a).
The blue cells are of $Z$-type, and the red ones are of $X$-type.
Within each blue cell we have eight 3-qubit $Z$ check operators\footnote{These are known as ``gauge operators'' in the literature. 
We use the name ``checks'' or ``check operators'', for the moment.
We will see that they will correspond to gauge fluxes after mapping to a gauge theory.}  $\bigtriangleup_Z \coloneqq Z_i Z_j Z_k$, which can be partitioned into two groups of four non-overlapping checks, which we color blue (B) and green (G), see Fig.~\ref{fig:lattice}(b).
Similarly, within each red cell, we have eight 3-qubit $X$ check operators, with colors red (R) and yellow (Y).
The colors are chosen such that the blue checks commute with yellow checks, and red checks commute with green checks.

One can form a 12-qubit operator by taking the product of all blue checks in a cell. This is equal to the product of all green checks in the same cell, see Fig.~\ref{fig:lattice}(c). The same constraint holds between the product of all red and yellow checks in a cell. 
\begin{subequations}
\label{eq:kinetic_gauss_laws}
\begin{align}
    \mbox{\mancube}_Z = \prod_{\text{blue } \bigtriangleup_Z \ \subset \ \mbox{\mancube}_Z} \bigtriangleup_Z 
    = \prod_{\text{green } \bigtriangleup_Z \ \subset \ \mbox{\mancube}_Z} \bigtriangleup_Z, \\
    \mbox{\mancube}_X = \prod_{\text{red } \bigtriangleup_X \ \subset \ \mbox{\mancube}_X} \bigtriangleup_X
    = \prod_{\text{yellow } \bigtriangleup_X \ \subset \ \mbox{\mancube}_X} \bigtriangleup_X.
\end{align}
\end{subequations}
We refer to the resulting 12-qubit operators as the ``stabilizers''. Notice that the stabilizers commute with all check operators.

The simplest and most natural Hamiltonian we can construct from the checks and stabilizers is the following~\cite{bombin2015gaugecolorcode, bombin2015singleshot, preskill2017qec},
\begin{align}
\label{eq:Hamiltonian_J_K}
    H_{\rm STC} =& - \frac{J_Z}{2} \sum_{\bigtriangleup_Z} \bigtriangleup_Z - \frac{J_X}{2} \sum_{\bigtriangleup_X} \bigtriangleup_X \nn
    & 
    - \frac{K_Z}{2} \sum_{\mbox{\mancube}_Z} \mbox{\mancube}_Z
    - \frac{K_X}{2} 
    \sum_{\mbox{\mancube}_X} \mbox{\mancube}_X.
\end{align}
Here we take $J_Z > 0$ and $J_X > 0$. 
For the moment, we assume translational symmetry for simplicity, so that $J_Z$ and $J_X$ do not depend on the locations or colors of the checks. We say a cell hosts an $e,m$ charge when the corresponding cube stabilizer obeys $\mbox{\mancube}_{Z,X} = -1$ respectively. 
The last two terms with $K_{Z,X} > 0$ are introduced to associate a gap to each $e$ or $m$ particle so that stabilizer violations are energetically suppressed.
Since the stabilizer terms commute with all the checks, the low energy subspace of $H$ is free of point charges regardless of the values of $J_{X,Z}$.
Charges cannot be generated by the Hamiltonian dynamics, but must be put in by hand or be introduced by a thermal bath.

Although the STC Hamiltonian in Eq.~\eqref{eq:Hamiltonian_J_K} is simple, it is not the most general one.
One can imagine a Hamiltonian in which some or all of the couplings are negative rather than positive.
However, if we want the low energy physics of the Hamiltonian to be free of errors, it is natural to suppress the errors energetically by setting the coupling strengths to be all positive.
One can also include all local terms generated by products of the check operators, and associate random couplings to them.
Addressing the thermal stability of the most general Hamiltonian is beyond the scope of this work.

\subsection{Limiting cases: 3D toric code topological order \label{sec:TC_limit}}

First we consider the $J_X = 0$ limit. All terms in the Hamiltonian commute. The cube terms penalize stabilizer errors, and the remaining check terms penalize blue and green flux loops. The excitations of the model involve violations of these cube and/or check terms.

\begin{figure}[!ht]
    \centering
    \includegraphics[width=.45\textwidth]{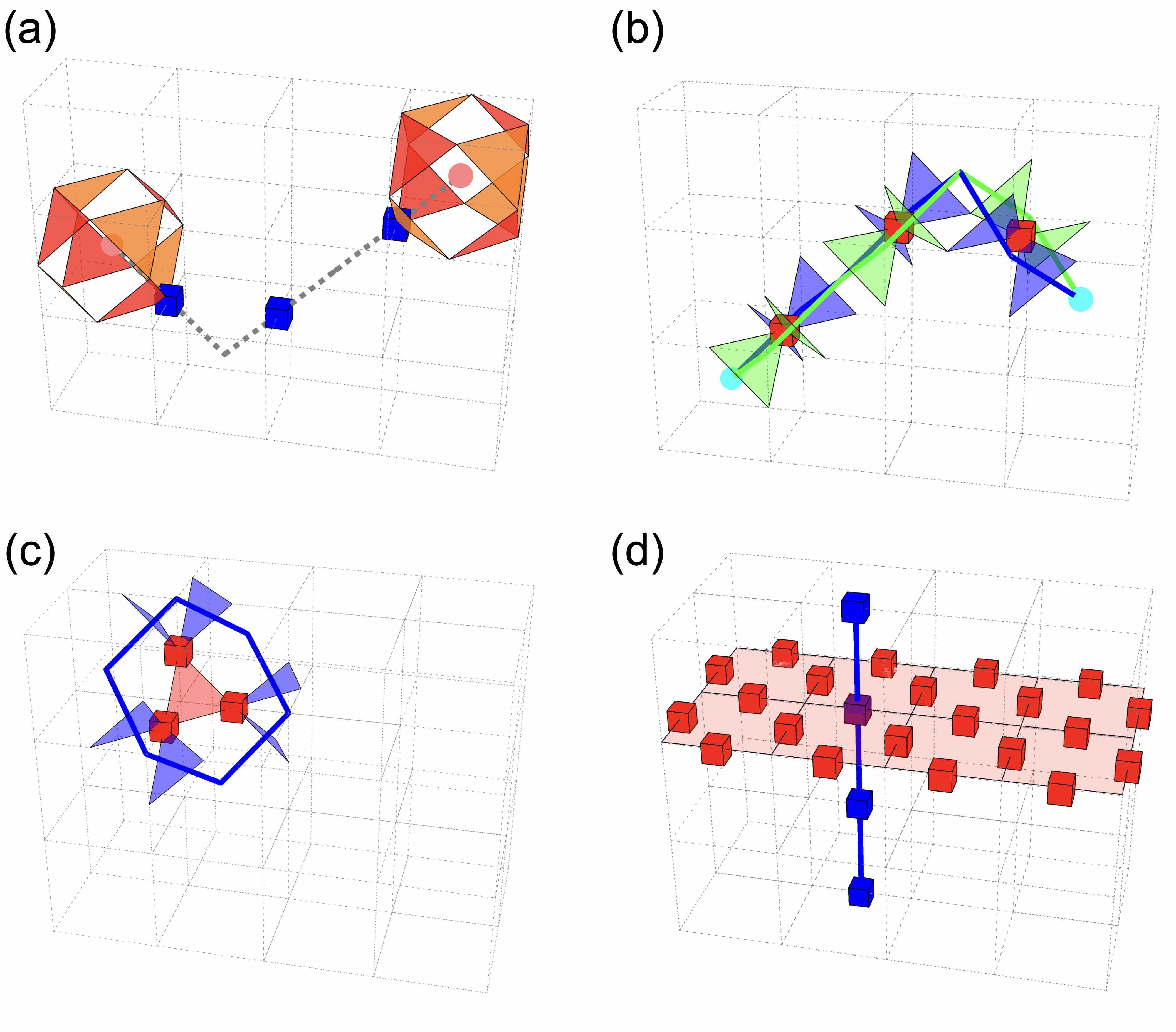}
    \caption{Point and loop excitations in the toric code limit $J_X = 0$.
    (a) A red check operator creates a closed loop of blue fluxes, which also costs an energy proportional to its length.
    (b) A pair of $m$ charges (represented by pink dots at the center of $X$ stabilizers) created by a string of $Z$ operators (represented by small blue cuboids).
    This excitation costs a finite energy, independent of the separation between the two $m$ charges.
    (c) With a string of $X$ operators (represented by small red cuboids), one creates a two $e$ charges (represented by cyan dots), as well as two strings of violated blue and green checks connecting the point charges.
    Together they form a closed flux loop.
    This excitation costs an energy that is proportional to the distance.
    This excitation can be composed with those in (a), so that the blue strings here do not necessarily track the $X$ Pauli operators.
    (d) 
    Logical operators of the toric code.
    Here the lattice is periodic in all three directions, and the logical operators are a closed loop of $Z$ operators, and a closed surface of $X$ operators, respectively.
    }
    \label{fig:TC_excitations}
\end{figure}

The blue/green check term violations can be created by acting with red/yellow checks. These create closed, blue/green loops with energy costs linear in their lengths. Consider, for example, a red $\bigtriangleup_X$ check, as shown in Fig.~\ref{fig:TC_excitations}(a).
It commutes with all green $\bigtriangleup_Z$ checks, as well as all $\mbox{\mancube}_Z$ stabilizers, but anticommutes with six blue $\bigtriangleup_Z$ checks, each sharing a qubit with it.
It thus creates a flux loop of length six, with a nonzero line tension $J_Z$.

There are two types of cube violations. Red/yellow cube violations are deconfined because they exude red/yellow fluxes tubes and these cost no energy as $J_X=0$.
As we show in Fig.~\ref{fig:TC_excitations}(b), such point excitations ($m$ charges) can be created by a line of Pauli $Z$ operators, and they each carry energy $K_X$.

Blue/green cube violations are associated with confined excitations ($e$ charges), because they are associated with blue/green flux tubes which have a linear-in-distance energy cost.
We can form these excitations as follows. A single-site Pauli $X$ operator violates four $\bigtriangleup_Z$ terms -- two of which are blue and the other two green -- as well as two $\mbox{\mancube}_Z$ terms.
When we have a string of Pauli $X$ operators, see Fig.~\ref{fig:TC_excitations}(c), 
they result in two $e$ charges connected by a blue electric flux and by a green electric flux; together they form a closed flux loop. Such two-color flux loops can be composed together to form larger flux loops, where the blue and green fluxes closely track each other. A flux loop of length $\ell$ created this way has energy $E(\ell) = J_Z \cdot \ell + 2 K_Z$, and thus has a nonzero line tension $J_Z$.

When put on a 3-torus, the model has 3 logical qubits, and an ($2^3$)-fold topological ground state degeneracy.
The logical operators are Pauli $Z$ operators on a non-contractible loop, or Pauli $X$ operators on a non-contractible membrane, as shown in Fig.~\ref{fig:TC_excitations}(d).

Thus, other than the fact that two-color flux loops excitations have an extra energy cost due to $e$ charges, the situation here is completely analogous to the  standard 3D toric code on a cubic lattice~\cite{KubicaVasmer}.
As we have discussed, $e$ charges are confined whereas $m$ charges are deconfined in this limit.

The other limit, when $J_Z = 0$ but $J_X > 0$, is similar, except that the roles between $X$ and $Z$ are now exchanged.
Moreover, the two limiting points -- namely $J_X = 0$ and $J_Z = 0$ -- should extend to stable phases with small $J_X$ or $J_Z$, respectively, since these points are equivalent to toric codes and are topologically ordered where all excitations are gapped~\cite{2010JMP....51i3512B, 2011CMaPh.307..609B}.
Assuming this is indeed the case,
the two phases -- namely $J_X / J_Z \ll 1$ and $J_X / J_Z \gg 1$ -- are clearly distinct, and should be separated by at least one phase transition.

It is clear that the phase diagram should be symmetric under $\lambda \to \lambda^{-1}$, where $\lambda = J_X / J_Z$.
Under this transformation the roles of $e$ and $m$ are exchanged, and it can be thought of as an electromagnetic duality.

\subsection{Mapping to lattice gauge theory in the pure gauge sector \label{sec:lattice_duality}}
We now study the phase diagram of the model for general nonzero values of $J_X$ and $J_Z$, where the Hamiltonian contains non-commuting terms. In particular, when comparing the elementary excitations in Fig.~\ref{fig:TC_excitations}, the nonzero $J_X$ terms now introduce a nonzero ``bare'' tension to the red and yellow fluxes (connecting $m$ charges, see Fig.~\ref{fig:TC_excitations}(b)).
We will focus on the ``matter free'' or ``pure gauge'' sector of the Hilbert space, where there are no background charges, i.e. cells where $\mbox{\mancube}_X = -1$ or $\mbox{\mancube}_Z = -1$.
The projector onto this subspace is given by
\begin{align}
    \Pi_{\rm gauge} = \prod_{\mbox{\mancube}_Z} \frac{1+\mbox{\mancube}_Z}{2} 
    \prod_{\mbox{\mancube}_X} \frac{1+\mbox{\mancube}_X}{2}.
\end{align}

We will also assume the lattice is periodic in all three dimensions, i.e., a 3-torus, as we are interested in the bulk phase in this section.

\begin{figure}[b]
    \centering
    \includegraphics[width=.45\textwidth]{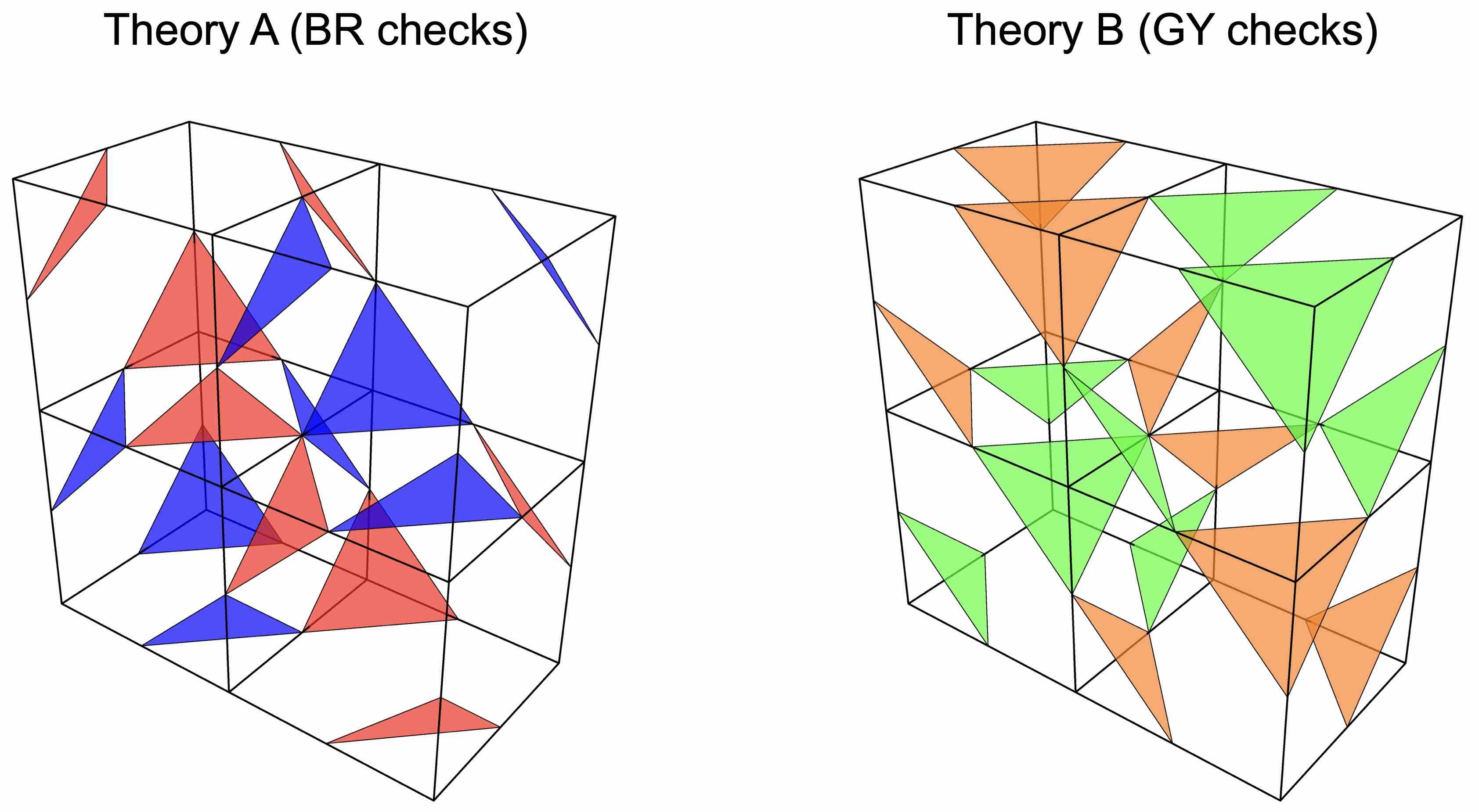}
    \caption{We divide the Hamiltonian into two set of operators, namely blue and red in one subset, and green and yellow checks in the other subset.
    Terms from different subsets commute.
    We can thus write the STC Hamiltonian in Eq.~\eqref{eq:Hamiltonian_J_K} as a sum of two mutually commuting ones, namely $H_{BR}$ and $H_{GY}$.
    }
    \label{fig:two_sublattice}
\end{figure}
As a first step in treating this noncommuting Hamiltonian, we notice that the check operators of the STC can be grouped into two disjoint subsets, and operators from different subsets commute with each other, see Fig.~\ref{fig:two_sublattice}. 
In one subset we have blue (B) and red (R) checks, and in the other we have green (G) and yellow (Y) checks.
As we can see, B and Y checks always overlap on an even number of sites, therefore they commute. Same can be said about G and R checks.
The two subsets of terms are only coupled through stabilizers (see Eq.~\eqref{eq:kinetic_gauss_laws}), and
in the absense of stabilizer violations the two subsets can be considered separately as two theories, which we denote as $H_{BR}$ and $H_{GY}$.
Within each theory, there are four non-overlapping check operators in each cubic cell.

\begin{figure}[!t]
    \centering
    \includegraphics[width=.45\textwidth]{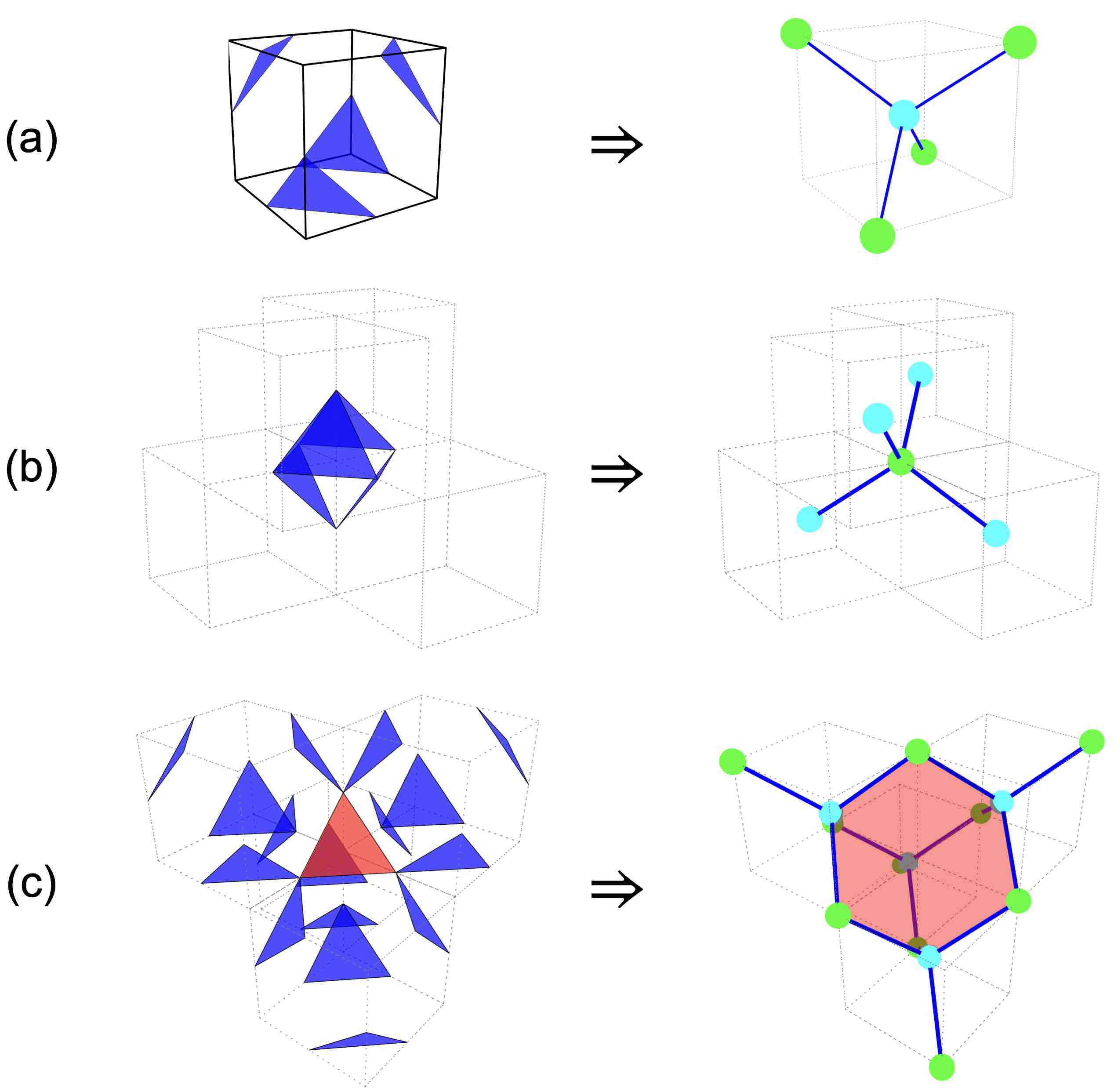}
    \caption{
    Mapping the STC Hamiltonian to a $\mb{Z}_2$ gauge theory on a diamond lattice.
    The vertices of the diamond lattice are cell centers (cyan) and cell corners (lime) of the original cubic lattice.
    (a) We identify each blue check with a dual spin $\tau_{ij}^x$ on each bond between a cyan and a lime vertex.
    These are to be interpreted as the electric flux.
    The stabilizer constraint thus becomes a lattice Gauss law, see Eq.~\eqref{eq:Z2_GaussLaw_cyan}.
    (b) Around each lime vertex there is an operator identity, namely the four blue checks shown in the figure should multiply to identity.
    This becomes another Gauss law Eq.~\eqref{eq:Z2_GaussLaw_lime}.
    Together with  Eq.~\eqref{eq:Z2_GaussLaw_cyan} we have the Gauss law everywhere on the lattice.
    (c) 
    Each red check flips six blue checks, and is associated with the lattice curl of the gauge flux, see Eq.~\eqref{eq:Z2_red_check_lattice_curl}.
    }
    \label{fig:face_to_bond}
\end{figure}

We now focus on $H_{BR}$ and describe a simplified representation.
We represent each $\bigtriangleup_Z$ check with a bond from the cell center to the cell corner (see Fig.~\ref{fig:face_to_bond}(a)), such that the bond is perpendicular to the triangular plaquette.
This way, we obtain a diamond lattice, where the vertices are cell centers (cyan) and cell corners (lime) from the original lattice.
We may formally define an ``effective spin'' on each bond for the corresponding $\bigtriangleup_Z$ check,
\begin{align}
\label{eq:Z2_plaquette_bond_duality}
    \tau^x_{ij} = \tau^x_{ji} = \bigtriangleup_Z.
\end{align}
With this new variable, the stabilizer constraint from the $K_Z \, \mbox{\mancube}_Z$ term can be interpreted as an ``electric'' lattice Gauss law (see Fig.~\ref{fig:face_to_bond}(a)),
\begin{align}
\label{eq:Z2_GaussLaw_cyan}
    \forall \text{ cyan } i, \quad
    \mbox{\mancube}_Z = \prod_{j \in n(i)} \tau^x_{ij} = 1.
\end{align}
Here $n(i)$ represents the vertices that are directly incident to $i$.
The operator $\tau^x_{ij}$ is thus an electric flux.
Another electric Gauss law constraint comes from an operator identity on each lime vertex, as can be seen from Fig.~\ref{fig:face_to_bond}(b),
\begin{align}
\label{eq:Z2_GaussLaw_lime}
    \forall \text{ lime } j,\quad \prod_{i \in n(j)} \tau^x_{ij} = 1.
\end{align}

We can also represent the $\bigtriangleup_X$ checks in terms of the new variable $\tau$.
As can be seen from Fig.~\ref{fig:face_to_bond}(c), each $\bigtriangleup_X$ check anticommutes with six $\bigtriangleup_Z$ checks, which form a hexagon on the diamond lattice.
Thus, we would expect
\begin{align}
\label{eq:Z2_red_check_lattice_curl}
    \bigtriangleup_X =&\  \prod_{\avg{ij} \in \hexagon}\tau^z_{ij}.
\end{align}
This is the magnetic flux.
As a consistency check, we see that the ``magnetic'' Gauss laws for the $X$ checks (i.e. the analogs of Eqs.~(\ref{eq:Z2_GaussLaw_cyan}, \ref{eq:Z2_GaussLaw_lime}) and  Fig.~\ref{fig:face_to_bond}(a, b) for red plaquettes) now become operator identities in terms of the $\tau$ spins.
Thus, imposing the ``magnetic'' Gauss laws (i.e. the constraint that all $\mbox{\mancube}_X = 1$) is crucial for this change of variables to go through.

On a 3-torus, this change of variables can be performed everywhere without running into subtleties.
In terms of the new variables, the Hamiltonian reads
\begin{align}
\label{eq:H_LGT}
    H^{\rm LGT} = - \frac{J_Z}{2} \sum_{\avg{ij}} \tau_{ij}^x - \frac{J_X}{2} \sum_{\hexagon} \prod_{\avg{ij} \in \hexagon}\tau^z_{ij},
\end{align}
with Gauss law constraints $\prod_{j \in n(i)} \tau^x_{ij} = 1$ on every vertex $j$.
The new Hamiltonian takes the familiar form of a $\mb{Z}_2$ lattice gauge theory (LGT) in three dimensions~\cite{fradkinsusskind1978}.
We note that a similar effective Hamiltonian has been obtained before by Kubica and Yoshida for the 3D gauge color code, through a different approach called ``ungauging''~\cite{kubica2018ungauging}.

The procedure we just described is commonly known as a ``lattice duality transformation'' in statistical mechanics.
In performing this transformation,
we made use of the observation that all stabilizer violations (i.e. $e$ and $m$ particles) are energetically suppressed at zero temperature, and therefore the system is in the ``pure gauge'' sector.
Therefore, at zero temperature,
the duality transformation is correct, in the sense that the dual variables and the original variables obey the same algebra.
Consequently, $H^{\rm LGT}$ and $H_{BG}$ have the same bulk spectrum when the energy scale is below $K_{X,Z}$, and their zero temperature phase diagram as tuned by $J_X/J_Z$ should be identical.
Formally, there exists a 
locality-preserving Clifford unitary $U$ such that
\begin{align}
    H^{\rm LGT} = U \Pi_{\rm gauge} H_{BR} \Pi_{\rm gauge} U^\dg.
\end{align}
A similar result holds for $H_{GY}$.

\begin{figure}[t]
    \centering
    \includegraphics[width=.48\textwidth]{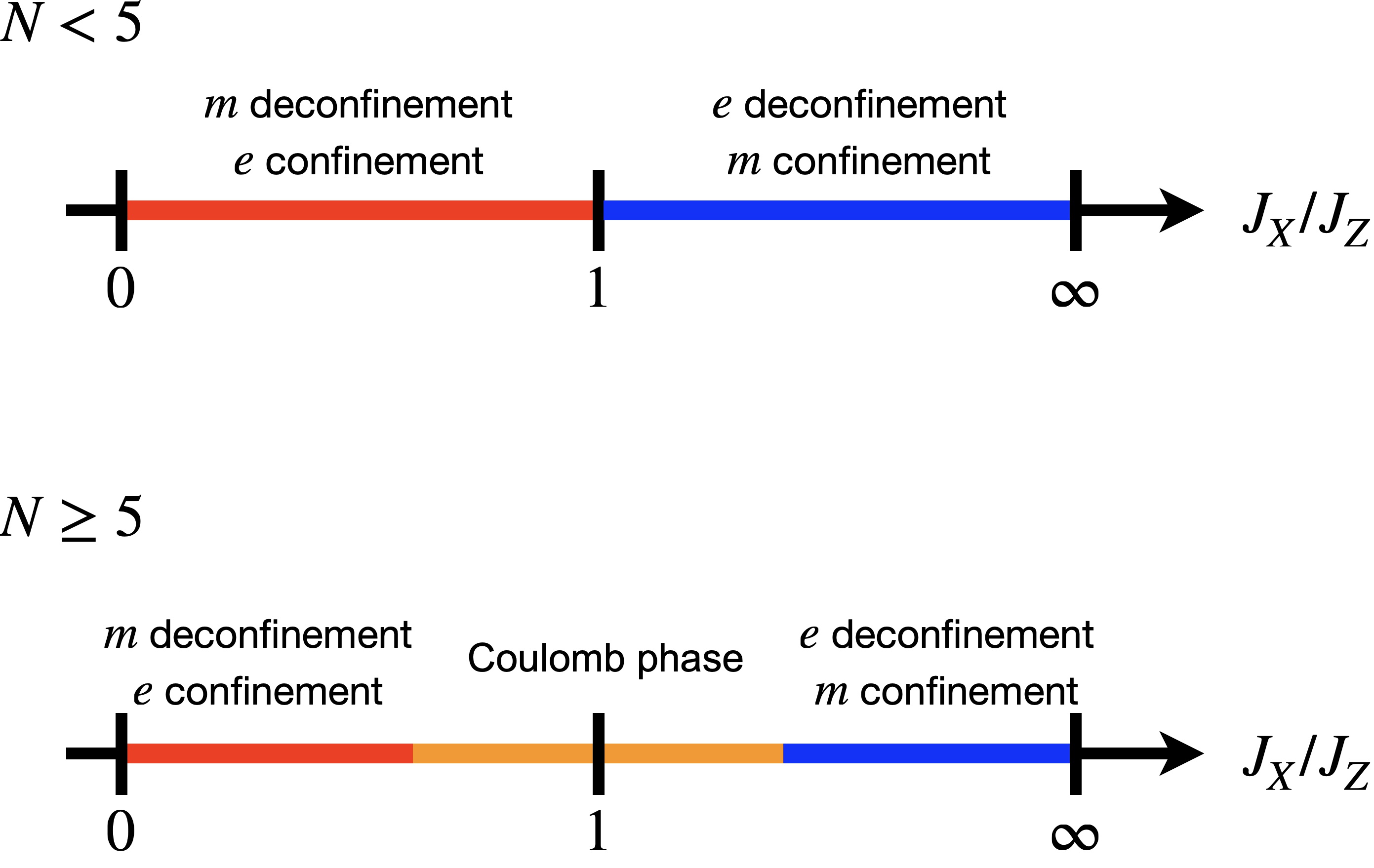}
    \caption{(a) Phase diagram of the $\mb{Z}_2$ LGT, Eq.~\eqref{eq:H_LGT}.
    There are two phases of this model, dual to each other, and are related by a first-order phase transition at $J_X/J_Z = 1$.
    (b)
    We can generalize the subsystem code to $N$-dimensional qudits, and the effective Hamiltonian is a $\mb{Z}_N$ gauge theory.
    When $N \ge 5$, there is an intermediate Coulomb phase near $J_X/J_Z = 1$.
    See Appendix~\ref{sec:ZN_generalization} for details.
    }
    \label{fig:ZN_phase_diagram}
\end{figure}

The 3D $\mb{Z}_2$ LGT on a \textit{cubic} lattice is well studied:
it has an ``electromagnetic'' self-duality~\cite{wegner1971, fradkinsusskind1978}, and its phase diagram has only two phases (dual to each other), separated by a first-order phase transition at the self-dual point~\cite{Creutz1979Ising}, see Fig.~\ref{fig:ZN_phase_diagram}(a).
This duality is most easily understood in the 4D classical  $\mb{Z}_2$ gauge theory on the 4D \textit{hypercubic} lattice that describes the Euclidean time path integral of $H^{\rm LGT}$, where plaquettes are dual to plaquettes.
In Appendix~\ref{sec:classical_MC} we obtain the same phase diagram for $H^{\rm LGT}$ on the diamond lattice by mapping to the 4D classical  $\mb{Z}_2$ gauge theory and subsequently performing classical Monte Carlo on the classical theory.
In our case, the electromagnetic duality is evident from the quantum Hamiltonian $H$ due to the symmetric role between $X$ and $Z$ checks.\footnote{The self-duality can also be seen from the level of the effective Hamiltonian $H^{\rm LGT}$.
If we associate a $\sigma$ spin to each $X$ check rather than each $Z$ check, we would get a Hamiltonian of the similar form as  Eq.~\eqref{eq:H_LGT}, but on the dual lattice of the diamond lattice in Fig.~\ref{fig:face_to_bond}(c), which is again a diamond lattice,
\begin{align}
    H^{\rm LGT} = - \frac{J_X}{2} \sum_{\avg{ij}} \sigma_{ij}^x - \frac{J_Z}{2} \sum_{\hexagon} \prod_{\avg{ij} \in \hexagon}\sigma^z_{ij}.
\end{align}
}

From the phase diagram of the 4D classical gauge theory we see that the two toric-code like topologically ordered phases, at $J_X / J_Z > 1$ and $J_X / J_Z < 1$ respectively,  are the only possible phases of the quantum Hamiltonian, and both have deconfined point charges ($e$ and $m$ charges in the two phases, respectively).
The phase diagram can also be obtained by perturbatively evaluating the energy cost of a pair of test charges in the quantum Hamiltonian $H^{\rm LGT}$.
As long as $J_X / J_Z < 1$, an eigenstate with two $m$ charges has a energy cost that is a convergent series expansion in $J_X/J_Z$, but is independent of their spatial separation.
In other words, a small bare string tension $J_X$ is not sufficient to confine the charges.

As in the 3D toric code, there is no topological order in equilibrium
at any finite temperature.
If we consider a nonequilibrium setting where we first prepare a error-free state and put it at finite temperature,
a finite density of deconfined point excitations will be introduced by thermal noise, and they can separate form one another at no energy cost. As a result, logical errors will occur in constant time~\cite{brown2016selfcorrectingRMP}. 

In Appendix~\ref{sec:ZN_generalization}, 
we generalize the qubit STC to qudit systems where each qudit has onsite dimension $N$, and the associated Hamiltonian is a $\mb{Z}_N$ gauge theory.
When $N$ is sufficiently large, an intermediate ``Coulomb'' phase is present~\cite{tHoof1978permanent, elitzur1979ZN, horn1979ZN, ukawa1980ZN}, see the phase diagram in Fig.~\ref{fig:ZN_phase_diagram}(b).
The Coulomb phase has gapless ``photon'' excitations and gapped, deconfined $e$ and $m$ charges.

\section{Bulk confinement and boundary toric code \label{sec:boundary_surface_code}}

We have been discussing the case where all the $\bigtriangleup_Z$ have the same coupling strengths $J_Z$ (and similarly for all the $\bigtriangleup_X$).
Recall that the two gauge theories (one for each subset in Fig.~\ref{fig:two_sublattice}) are decoupled except that they share point excitations (and therefore point excitations are sources of fluxes in both theories, see Eq.~\eqref{eq:kinetic_gauss_laws} 
and Fig.~\ref{fig:TC_excitations}(c)), but so far this kinetic constraint has not entered our discussion of the phase diagram.
The phases we found so far are no different from the 3D toric code.
Here we consider assigning different coupling strengths to the two theories, while maintaining the translational invariance within each of them.
As we will see, in this regime the kinetic constraint plays an important role, and we can understand certain aspects of the subsystem code from the gauge theory.

\begin{figure}[!b]
    \centering
    \includegraphics[width=.48\textwidth]{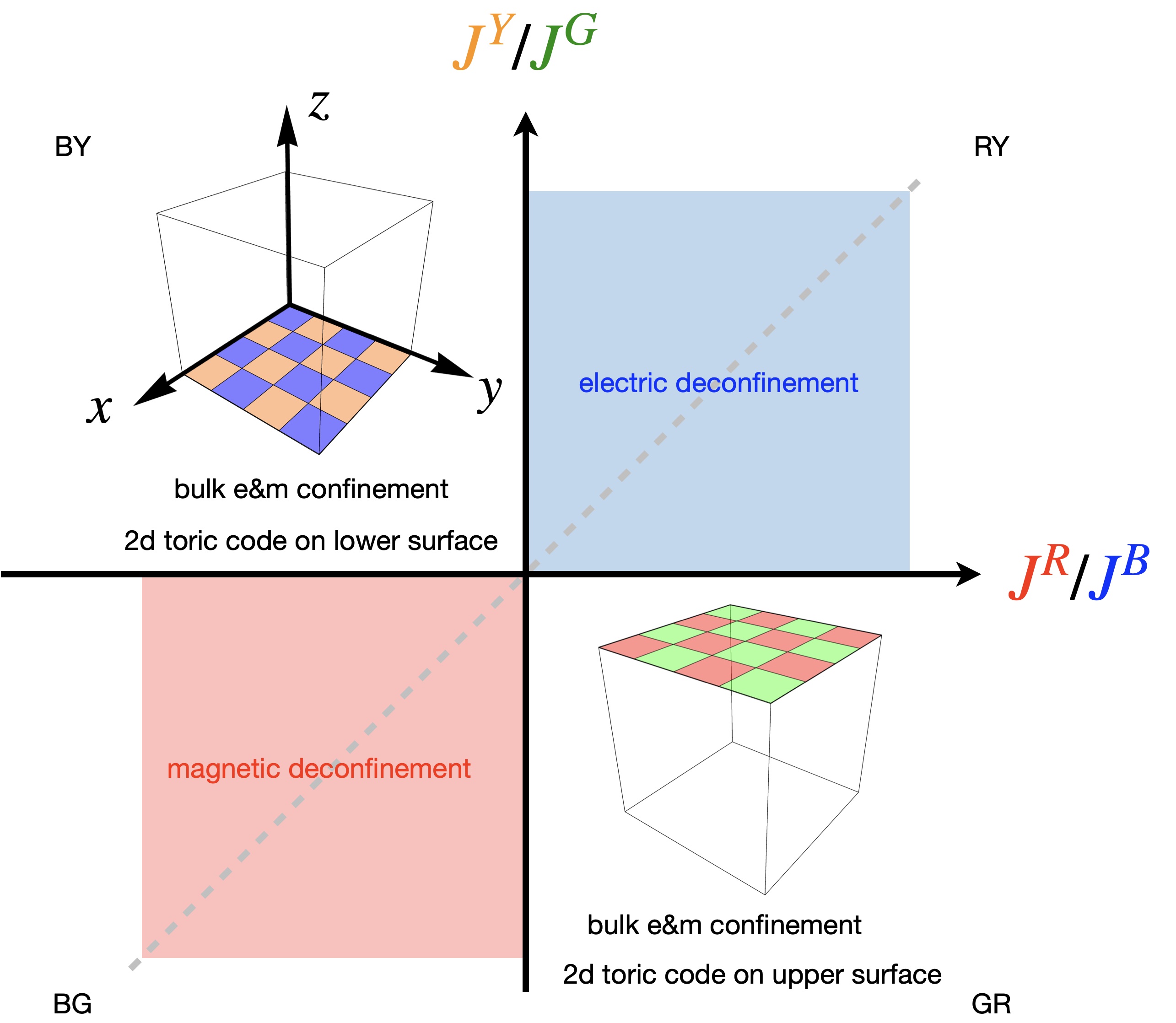}
    \caption{Phase diagram of the STC Hamiltonian when we associate different coupling strengths in the two sublattices.
    The dashed line along the diagonal corresponds to the phase diagram  discussed in Fig.~\ref{fig:ZN_phase_diagram}(a), where we fixed $J^B = J^G = J_Z$ and $J^R = J^Y = J_X$.
    See main text.
    }
    \label{fig:offdiagonal_phase_diagram}
\end{figure}

\begin{figure*}
    \centering
    \includegraphics[width=.9\textwidth]{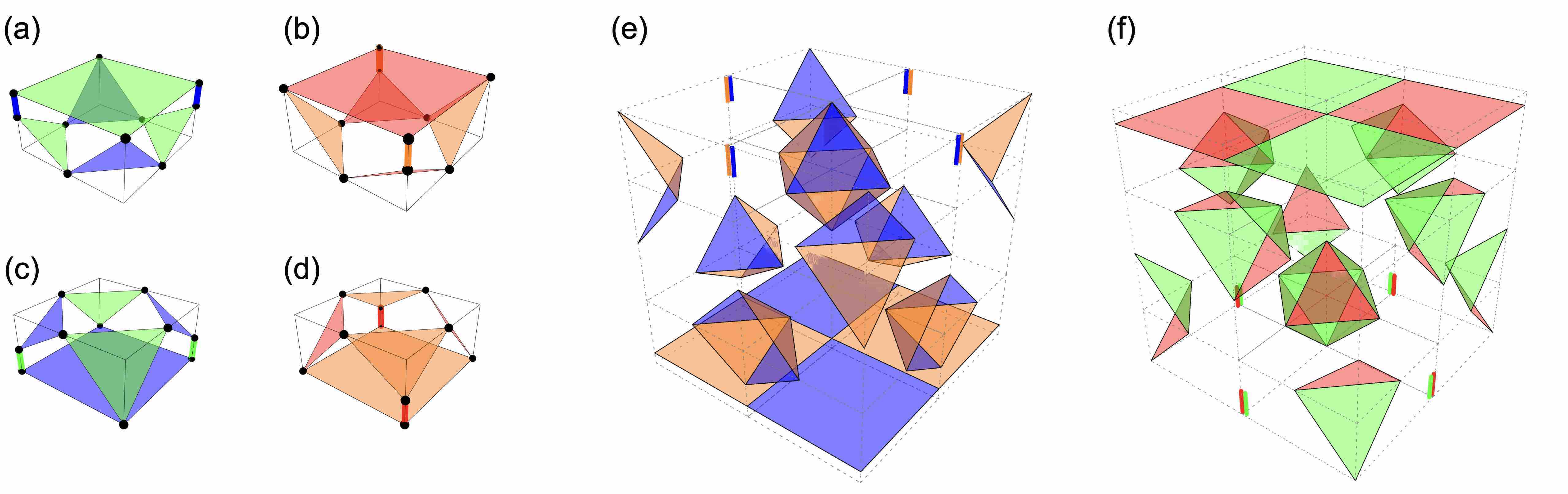}
    \caption{(a-d) Boundary conditions for the STC as constructed in Ref.~\cite{KubicaVasmer}.
    With this boundary condition, we illustrate the nonzero terms (e) in the limit $J^G = J^R = 0$, and (f) in the limit $J^B = J^Y = 0$.
    In (e) we have decoupled local systems in the bulk, and a toric code on the lower boundary.
    The picture is similar in (f), where the toric code is now on the upper boundary.
    }
    \label{fig:boundary_details}
\end{figure*}

To simplify the notation, we denote the four coupling strengths with their corresponding color, namely $J^{BR}_Z = J^B$, $J^{BR}_X = J^R$, $J^{GY}_Z = J^G$, and $J^{GY}_X = J^Y$.
The phase diagram is two dimensional, with axes $J^R / J^B$ and $J^Y / J^G$, see Fig.~\ref{fig:offdiagonal_phase_diagram}.
The regime we considered previously is then along the diagonal of this two-dimensional phase diagram.

Consider the upper-left quadrant of Fig.~\ref{fig:offdiagonal_phase_diagram}, where $J^R / J^B < 1$ but $J^Y / J^G > 1$.
We refer to this as the ``BY'' phase, after the dominant blue and yellow check terms.
$H_{BR}$ is now in the phase with magnetic deconfinement and electric confinement, whereas $H_{GY}$ is in the dual phase.
A pair of test $e$ charges is shared between theories $H_{BR}$ and $H_{GY}$, and as required by the kinetic constraint (and as we have seen in Fig.~\ref{fig:TC_excitations}), they are connected by fluxes in both theories, one of blue color and the other green.
Its energy is simply the sum over the two theories, and is therefore dominated by the (blue) electric flux tension in $H_{BR}$, which is confining.
Similarly, the energy of a pair of test $m$ charges is dominated by the (yellow) magnetic flux tension in $H_{GY}$, thus also confined.
As a consequence of the kinetic constraint, 
in this part of the phase diagram there are no deconfined bulk excitations.

The bulk confinement is also easily seen on the lattice level.
First take the limit $J^R = J^G = 0$.
As we illustrate in Fig.~\ref{fig:boundary_details}(e), the nonzero check terms in the bulk are now commuting and disconnected, and they form unfrustrated and decoupled local systems (blue-yellow octahedra in Fig.~\ref{fig:boundary_details}(e)).
When put on a closed manifold, the system has a unique global ground state 
that is the product of the local ground states.\footnote{In the bulk, these are stabilizer states that are each the unique codestate of a [[6,0,3]] quantum error correcting code.}
Concomitant with this, there should be no deconfined charges.
Indeed, a pair of point excitations will necessarily excite a trail of octahedra, with an energy linear in their separation, thus confined.
This picture is unchanged as long as $J^R / J^B < 1$ and $J^Y / J^G > 1$,\footnote{As we learn from the phase diagram in Fig.~\ref{fig:ZN_phase_diagram}, the effects of nonzero $J^R$ and $J^G$ are perturbative, and only introduce a finite correlation length.} and the entire upper-left quadrant is in the same  bulk BY phase.
The lower-right quadrant (GR phase) can be discussed similarly.

Without ground state degeneracy, the system cannot be used as a topological code.
Indeed, for the 3D STC there are no logical qubits when the lattice is periodic in all directions~\cite{KubicaVasmer}.

We now turn to lattices with a boundary, which is more interesting.
There is no canonical boundary condition, and we adopt a construction by Kubica and Vasmer~\cite{KubicaVasmer}, where such a boundary condition leads to a nonzero number of logical qubits of the subsystem code.

In the rest of this section, we first give a microscopic description of the boundary condition from Ref.~\cite{KubicaVasmer} in Sec.~\ref{sec:bc_microscopics}; then we provide a more general description in Sec.~\ref{sec:bc_fluxes} in terms of condensation of gauge fluxes.
In Sec.~\ref{sec:bare_logicals} we describe the bare logical operator of the code, and discuss how they come from those of the boundary toric codes. 

Furthermore, in Appendix~\ref{sec:summary_excitations}, we provide a complete table of excitations and operator identities that are useful for discussions of this section.
In Appendix~\ref{sec:bc_further_discussions}, we provide physical motivations for the (somewhat complicated) boundary condition in Fig.~\ref{fig:boundary_details}, and discuss a few different ones that fail to give logical qubits to the code.

\subsection{Microscopics of the boundary condition \label{sec:bc_microscopics}}

We take the lattice to be periodic in $x$ and $y$ directions, and open in the $z$ direction, so that lattice topology is $\mathbb{T}^2\times [0, 1]$.
Let the lattice have $L_x \times L_y \times L_z$ cells.
Near the upper and lower boundaries, the 12-qubit stabilizers are replaced by 10-qubit stabilizers, see Fig.~\ref{fig:boundary_details} for an example where $L_x = L_y = 2$, $L_z = 3$.
In particular, Fig.~\ref{fig:boundary_details}(a,b) represent $Z$ and $X$ stabilizers on the upper boundary, and Fig.~\ref{fig:boundary_details}(c,d) represent $Z$ and $X$ stabilizers on the lower boundary.
Each of these cells has seven checks.
In addition to 3-qubit checks, there are now also 2-qubit and 4-qubit checks.
Each 2-qubit check is shared between two cells connected in the diagonal direction.

As we show in Fig.~\ref{fig:boundary_details}(e), when $J^R = J^G = 0$ (deep in the BY phase of Fig.~\ref{fig:offdiagonal_phase_diagram}), the terms in the Hamiltonian all commute.
In the bulk and on the upper boundary we still have a product of decoupled local systems; but on the lower boundary we now have a 2D toric code.
Similarly, when $J^Y = J^B = 0$ (GR phase of Fig.~\ref{fig:offdiagonal_phase_diagram}), we have a 2D toric code on the upper boundary, see Fig.~\ref{fig:boundary_details}(f).
Each phase has deconfined anyon excitations, two logical qubits, and a four-fold ground state degeneracy --- all coming from the topological order of the boundary toric code.

\begin{figure}[t]
    \centering
    \includegraphics[width=.25\textwidth]{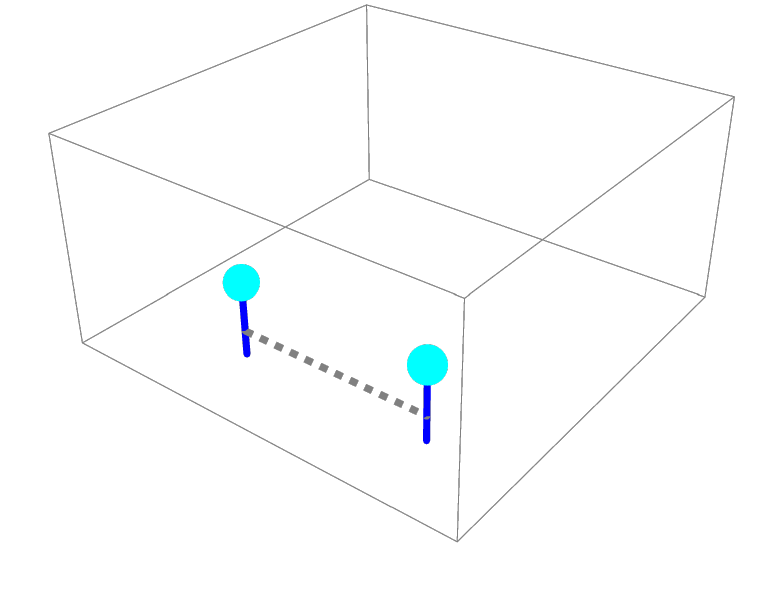}
    \caption{A pair of $e$ charges that are finite distance away from the boundary are still deconfined in the BY phase.
    The lowest energy configuration, shown here, has an energy proportional to the length of the blue flux strings, but does not grow with the distance, due to deconfinement of boundary fluxes (i.e. anyons) in the boundary toric code.}
    \label{fig:shortcut}
\end{figure}

As an aside, we note that the deconfined charges of the boundary 2D toric code are effectively pinned to live near the boundary by a linearly growing potential.
As we show in Fig.~\ref{fig:shortcut}, the two $e$ charges each has a gap proportional to their distance to the boundary, and as long as the gap remains a constant they remain deconfined.

\begin{figure}[t]
    \centering
    \includegraphics[width=0.45\textwidth]{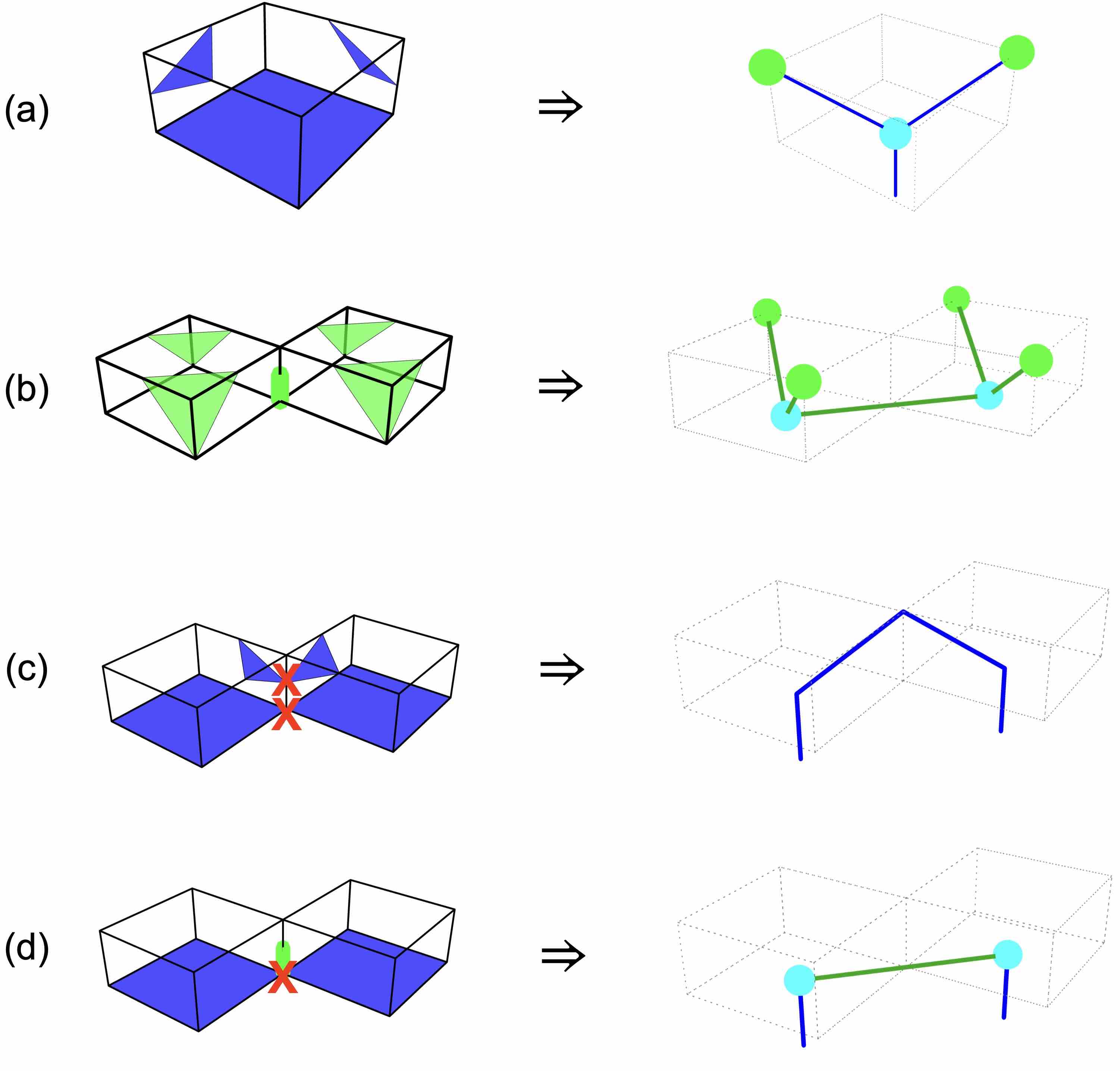}
    \caption{(a,b) Mapping the boundary checks to fluxes.
    The triangular blue and green checks are similar to Fig.~\ref{fig:face_to_bond}.
    In (a), the 4-qubit blue check on the lower boundary maps to a ``boundary flux'', or an anyon of the boundary toric code.
    In (b), the 2-qubit green check is mapped to an in-plane green flux connecting neighboring $e$ charges.
    (c,d) A blue boundary flux (i.e. blue anyon) can be created either by a 2-qubit red check operator as in (c), or by a single-qubit Pauli $X$ as in (d).
    The former (c) creates an open blue flux, and the latter (d) creates two $e$ charges connected by a green flux, and each carrying a blue flux on the boundary.
    }
    \label{fig:boundary_check_to_flux}
\end{figure}

\begin{figure*}[t]
    \centering
    \includegraphics[width=1.0\textwidth]{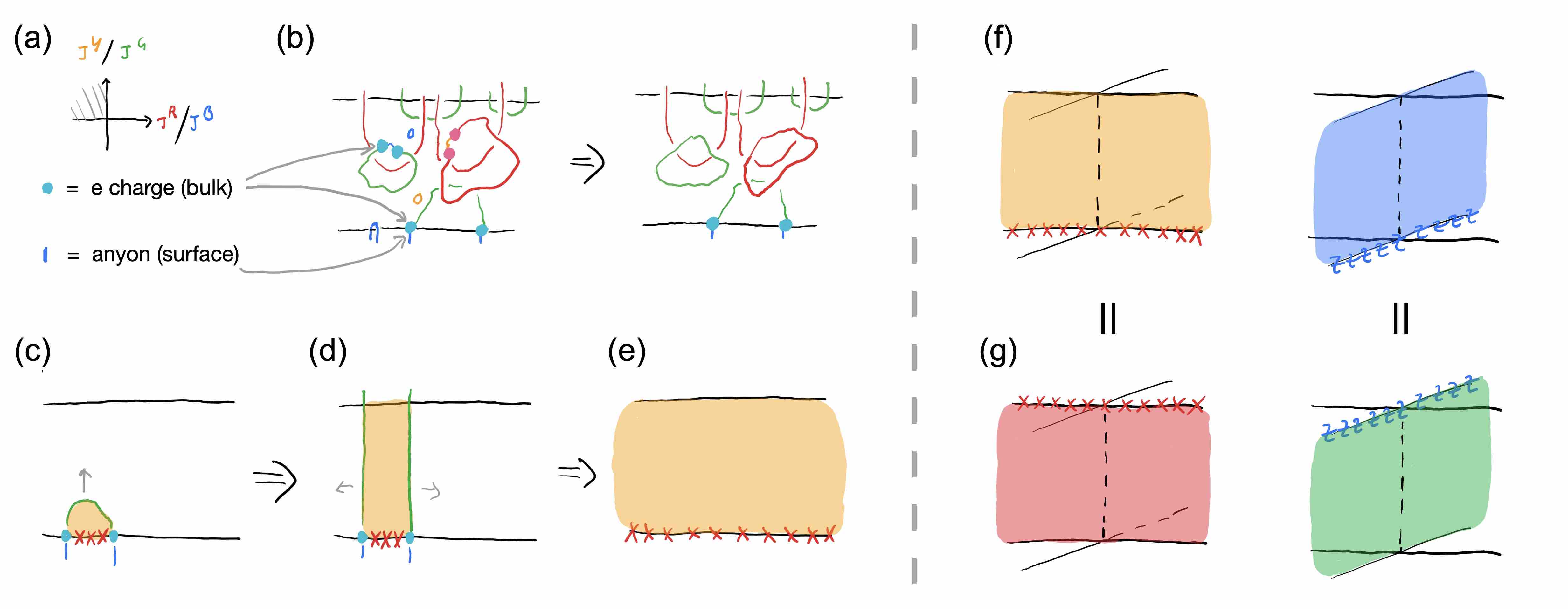}
    \caption{(a,b) In the BY phase at long distances, the system is a condensate of red and green flux loops (in the bulk) or half flux loops (near the upper boundary).
    Blue and yellow fluxes have a large tension and can be neglected at long distances.
    Thus, the bulk is completely confined, as point charges will necessarily be connected by blue or yellow fluxes.
    (c) The elementary excitations are ``anyon-charge-flux'' composites, namely pairs of $e$ charges near the lower boundary, each carrying a blue boundary flux, and themselves connected by a green flux, see also Fig.~\ref{fig:boundary_check_to_flux}(d).
    The two charges can be separated by a string of Pauli $X$s.
    This configuration has a constant energy and is deconfined. 
    (d) The green flux in ``anyon-charge-flux'' composite can be moved by green flux loops from the condensate.
    In particular, we can bring it all the way up to the upper boundary, where the green fluxes can leave the system without costing energy.
    (e) Finally, we can bring the pair apart so that they trace out a nontrivial surface of the lattice, before they annihilate in pairs.
    The operator for this process does not leave any syndromes, and is a bare logical operator of the code.
    (f) The bare logical operator constructed in (c-e) is the product a logical operator of the boundary toric code and a surface of yellow checks.
    Its canonical conjugate (right panel) can be similarly constructed as a product of the conjugate logical operator of the boundary toric code and a surface of blue checks.
    The anticommutation between the two bare logical operators follows from the boundary toric code, whereas the yellow and blue checks commute in the bulk.
    (g) In the GR phase, the same bare logical operators can also be created, although through a different process.
    Accordingly, they aquire a different representation.
    The equality can be checked with operator identities discussed in Appendix~\ref{sec:summary_excitations}.
    }
    \label{fig:bare_logicals}
\end{figure*}

\subsection{Boundary anyons at endpoints of bulk fluxes \label{sec:bc_fluxes}}

It is useful to translate the boundary conditions into the language of fluxes, as we did in the bulk.
In Fig.~\ref{fig:boundary_check_to_flux} we illustrate the case of a $Z$ cell near the lower boundary (i.e., the cell in Fig.~\ref{fig:boundary_details}(c)); other cells should follow similarly.
Notably, some gauge fluxes can now terminate on the boundary.
As shown in Fig.~\ref{fig:boundary_check_to_flux}(a), in the absence of $e$ charges the Gauss law on the blue checks requires that a blue flux coming from the bulk can either go back into the bulk, or terminate on the boundary.
In the latter case, we identify such boundary fluxes as \textit{anyons} of the boundary 2D toric code.\footnote{\label{fn:anyon_vs_charge}We emphasize that this ``blue anyon'' is a point excitation of the boundary toric code, and should be distinguished from point excitations ($e$ and $m$ charges) in the bulk.
In all our figures, we represent the blue anyon with a blue flux sticking out of the boundary, whereas an $e$ charge is always represented with a cyan dot, as consistent with our convention in Figs.~\ref{fig:TC_excitations}, \ref{fig:face_to_bond}.}

The identification of 3-qubit green checks are similar as before, and the 2-qubit green checks are identified as an in-plane green flux connecting neighboring $e$ charges, see Fig.~\ref{fig:boundary_check_to_flux}(b).
Importantly, green fluxes cannot get out of this boundary.

Let us focus on the BY phase $J^R / J^B < 1$, $J^Y / J^G > 1$ (upper-left quadrant of Fig.~\ref{fig:offdiagonal_phase_diagram}).
There are two ways of creating blue anyons on the lower boundary.
\begin{enumerate}[i)]
    \item 
The first is to create an open string of blue fluxes that terminates on the boundary, without violating any stabilizers (thus do not create any $e$ charges). 
In Fig.~\ref{fig:boundary_check_to_flux}(c) we illustrate one example of this excitation, where the creation operator is the two-qubit red check in Fig.~\ref{fig:boundary_details}(d), containing two Pauli $X$s.
In fact, any red check in Fig.~\ref{fig:boundary_details}(d) at the lower boundary -- whether they act on 2 or 3 qubits -- creates this type of excitation, as it commutes with stabilizers (thus does not create $e$ charges) but anticommutes with blue checks (thus creates blue fluxes).
This type of configuration can be composed to create a long open blue flux string, thereby  seperating the two anyons at the endpoints.
When we do so, the excitation costs an energy linear in the separation, coming from the large string tension of the blue fluxes.
\item
The second is to create a pair of $e$ charges on the boundary at a finite energy cost, so that the boundary blue flux gets attached to green fluxes through the $e$ charge.
In Fig.~\ref{fig:boundary_check_to_flux}(d) we illustrate one example of this excitation, where the creation operator is a single-qubit Pauli $X$ operator.
Importantly, now the two $e$ charges need not be connected by blue checks in the bulk, and they become deconfined at the lower boundary.\footnote{In fact, due to our discussion in Fig.~\ref{fig:shortcut}, a pair of $e$ charges remain deconfined as long as they are a finite distance away from the lower boundary.} 
(The creation operator is a string of single-site Pauli $X$ operators.)
The case for the yellow fluxes on the lower boundary is similar.
\end{enumerate}

On the upper boundary, the allowed boundary fluxes are green and red.
They can be created by terminating bulk open flux strings of green/red color on the boundary, in a manner similar to item i) in the previous paragraph.
In the BY phase $J^R / J^B < 1$, $J^Y / J^G > 1$, the boundary green and red fluxes are gapless, and can be created at no energy cost.
The ground state is thus a condensate of green and red open flux loops terminating on the upper boundary, as well as green and red closed flux loops in the bulk.

After translating the excitations into gauge fluxes, we can describe the BY phase in a cartoon picture, see Fig.~\ref{fig:bare_logicals}(a,b), where we leave out all lattice details.
At long length scales, the theory flows to the fixed point described by the Hamiltonian where $J^G = J^R = 0$.
In this case, blue and yellow fluxes have a nonzero tension, and thus confined and negeligible in the bulk; they only appear on the lower boundary, as deconfined anyons (identified as boundary fluxes, see footnote~\ref{fn:anyon_vs_charge}).
We have a condensate of green and red flux loops (in the bulk) or half loops (near the upper boundary).
They must attach to blue or yellow anyons (via $e$ or $m$ charges) if they terminate on the lower boundary,
and can terminate anywhere on the upper boundary without costing energy.
The braiding statistics between blue and yellow anyons are entirely determined on the boundary; the red and green fluxes commute with each other in the bulk and do not affect the boundary toric code.

\subsection{Bare logical operators and their dependence on boundary conditions \label{sec:bare_logicals}}

The Hamiltonian model we consider has an interesting property.
There exists a set of operators -- so-called ``bare'' logical operators~\cite{bombin2015gaugecolorcode} --  which commute with our model for all choices of couplings.
It is important to understand such operators, due to their relevance to performing logical operations on encoded qubits.
However, it was not clear how such operators should be constructed, and why they should only exist with the correct boundary conditions.

Here, using the picture in Fig.~\ref{fig:bare_logicals}(a,b), we show how to obtain ``bare'' logical operators of the subsystem toric code in the BY phase.

Recall that the bare logical group is defined as 
\begin{align}
    \mc{L}^{\rm bare} \coloneqq Z(\mc{G}) / (\mc{G} \cap  Z(\mc{G}) ) = Z(\mc{G}) / \mc{S},
\end{align}
where $\mc{G} = \avg{\{\bigtriangleup_Z\}, \{\bigtriangleup_X\}}$ is the group generated by all of the triangular check operators,  $Z(\mc{G})$ is its centralizer, and $\mc{S}$ is the group of cube stabilizers.
(The equality holds because if two elements of $Z(\mc{G})$ differ by a product of $X$ checks, that product must be in $\mc{S}$ for both to commute with all $Z$ checks, since for each non-stabilizer $X$ check there is at least one $Z$ check that anticommutes with it, in our construction.)
A nontrivial bare logical operator thus commutes with all of the checks, but is not a product of checks, and is only defined up to stabilizers.
A bare logical operator is thus an exact symmetry of the Hamiltonian, and acts nontrivially within the degenerate ground space.

To find a bare logical operator, we follow a familiar approach, namely by looking at topologically nontrivial worldsheets of deconfined excitations.
Consider an operator that creates a pair of blue anyons in the boundary 2D toric code in the BY phase, which can be an arbitrary string of Pauli $X$ living on the boundary with the anyons at its endpoints, see Fig.~\ref{fig:bare_logicals}(c), which is a generalization of Fig.~\ref{fig:boundary_check_to_flux}(d).
As we have discussed, green fluxes near the boundary will necessarily be attached to the two isolated blue anyons by $e$ charges.
Since the bulk is a condensate of green flux loops, the shape of the green flux string is arbitrary, and different choices have identical actions on the logical space of the boundary code.
In other words, once we have an boundary operator that creates a pair of anyons and a green flux string, we are free to move it by creating or annihilating green flux loops (and half loops) in the condensate, as long as they are still attached to the boundary anyons.

In Fig.~\ref{fig:bare_logicals}(c-e), we illustrate a time evolution of the flux-charge-anyon composite of particular interest.
First, we bring the string all the way up to the upper boundary where they can be absorbed.
This results in two disjoint green strings emanating from the blue anyons running towards the upper boundary, see Fig.~\ref{fig:bare_logicals}(d).
Then, we can pull the two blue anyons apart and annihilate them after the two have completed a noncontractible loop, see Fig.~\ref{fig:bare_logicals}(e).
The green strings attached to them can similarly sweep out a noncontractible surface and annihilate.
Such a surface does not leave any syndromes, so an operator that creates it commutes with all elements in $\mc{G}$.
We have thus obtained an element of $\mathcal{L}^{\rm bare}$, which is a surface of yellow checks attached to the logical operator of the boundary toric code (namely a string of Pauli $X$s).\footnote{This logical operator of the boundary toric code \emph{without} an attached yellow surface is a \emph{dressed} logical operator for the STC, see Appendix~\ref{sec:dressed_logicals}.
The process outlined here mimicks the dominate logical error process in the BY phase when the STC Hamiltonian is put at finite temperature.}

A canonical conjugate to this bare logical operator can be constructed similarly by attaching a surface of blue checks to the conjugate logical operator of the boundary toric code (namely a string of Pauli $Z$s).
The conjugate pair is shown in Fig.~\ref{fig:bare_logicals}(f).
Both are supported on deformable membranes terminating the upper and lower boundaries.

It is quite unusual that the conjugate pair intersect on a line (rather than a point), but 
from our discussion it is clear that they should anticommute due to their contents on the boundary (whereas they commute in the bulk), which are nothing but logical operators of the boundary 2D toric code.
It is also clear that the procedure with which we obtain the bare logicals depends on rather general features, namely the condensation of green and red flux loops in the bulk and on the upper boundary, rather than lattice details.

Of course, the bare logical operators are defined by the code and is not merely a property of the BY phase.
The same bare logical operators can be obtained in the GR phase following a similar reasoning, as we show in Fig.~\ref{fig:bare_logicals}(g).
Here, it is appropriate to choose different representations of the same operator in different phases.
In the BG and RY phases (diagonal phases of Fig.~\ref{fig:offdiagonal_phase_diagram}, both equivalent to the 3D toric code), one of the logical operators can be reduced (by checks) to a logical operator supported on a line, while the other remains a surface logical operator of the 3D toric code.

A membrane that does not touch either boundary, such as one in the $xy$ plane (as illustrated in Fig.~\ref{fig:TC_excitations}(d)), cannot support a nontrivial logical operator in the subsystem code, as it is the product of all cube stabilizers bounded below (or of all bounded above), thus itself an element of $\mc{G}$.

\section{Check measurement sequences for single-shot error correction \label{sec:ssec}}

\begin{figure*}[ht]
    \includegraphics[width=.8\textwidth]{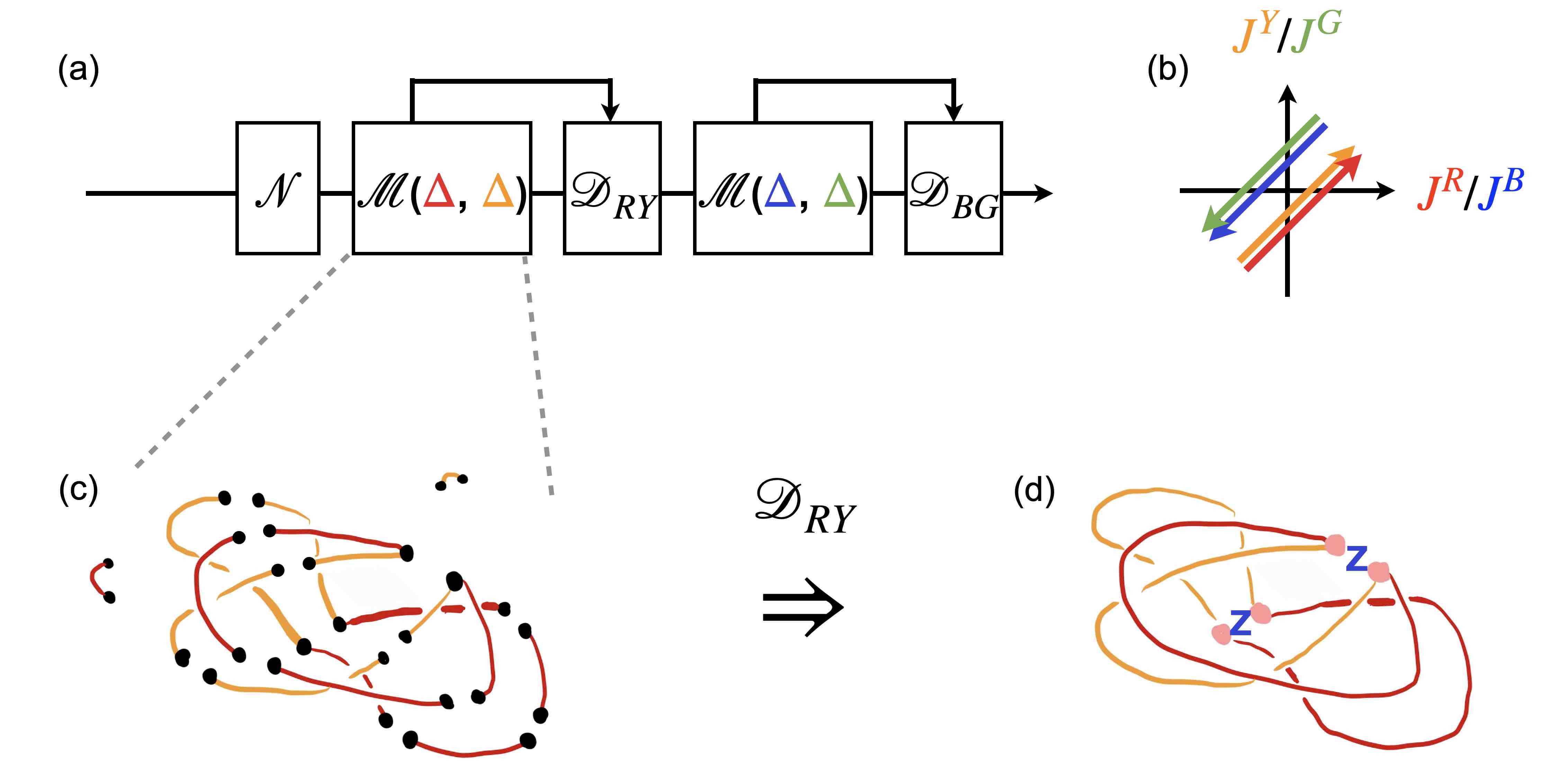}
    \caption{
    The standard way of doing single-shot error correction with the 3D STC, following Ref.~\cite{KubicaVasmer}.
    (a) Within each round, we measure all the $X$ checks (red and yellow) to decode and correct bit-flip errors, and then all the $Z$ checks (blue and green) to decode and correct phase errors.
    The decoders are denoted $\mathcal{D}_{RY}$ and $\mathcal{D}_{BG}$, respectively.
    The round is then repeated over and over again, and it was shown in Ref.~\cite{KubicaVasmer} that the logical error rate increases no faster than linear in time.
    Despite that the measurements are imperfect and we are only making $O(1)$ measurements of each check before decoding, the physical error threshold remains finite and roughly indepedent of time~\cite{KubicaVasmer}.
    (b) The state of the system immediately after RY measurements is an eigenstate of the RY checks (up to signs), and looks like a low-lying state of the 3D toric code with the stabilizer group generated by $\bigtriangleup_X$ (the RY checks) and $\mbox{\mancube}_Z$.
    After the BG measurements, the state instead looks like a low-lying state of the dual 3D toric code, as depicted in Fig.~\ref{fig:TC_excitations}.
    Thus, in a sense the system jumps between the two diagonal phases of phase diagram.
    (c) The results of RY measurements, where we highlight fluxes where the check measurement result is nontrivial.
    With measurement errors, the R and Y fluxes appear as closed loops with short missing segments.
    We also have short fluxes, which are ``false positive'' syndromes.
    (d) The decoder first repairs the fluxes for form closed loops, by performing a minimal weight perfect matching (MWPM) of the endpoints (black dots) of RY fluxes.
    Then stabilizer errors can be identified as cubes that emanate both R and Y fluxes (pink dots, as consistent with Fig.~\ref{fig:TC_excitations}(b)), and can be successfully distinguished from measurement errors when the error rates are small.
    The physical errors can further be paired, again with MWPM.
    }
    \label{fig:ssec}
\end{figure*}

\begin{figure*}
    \includegraphics[width=0.8\textwidth]{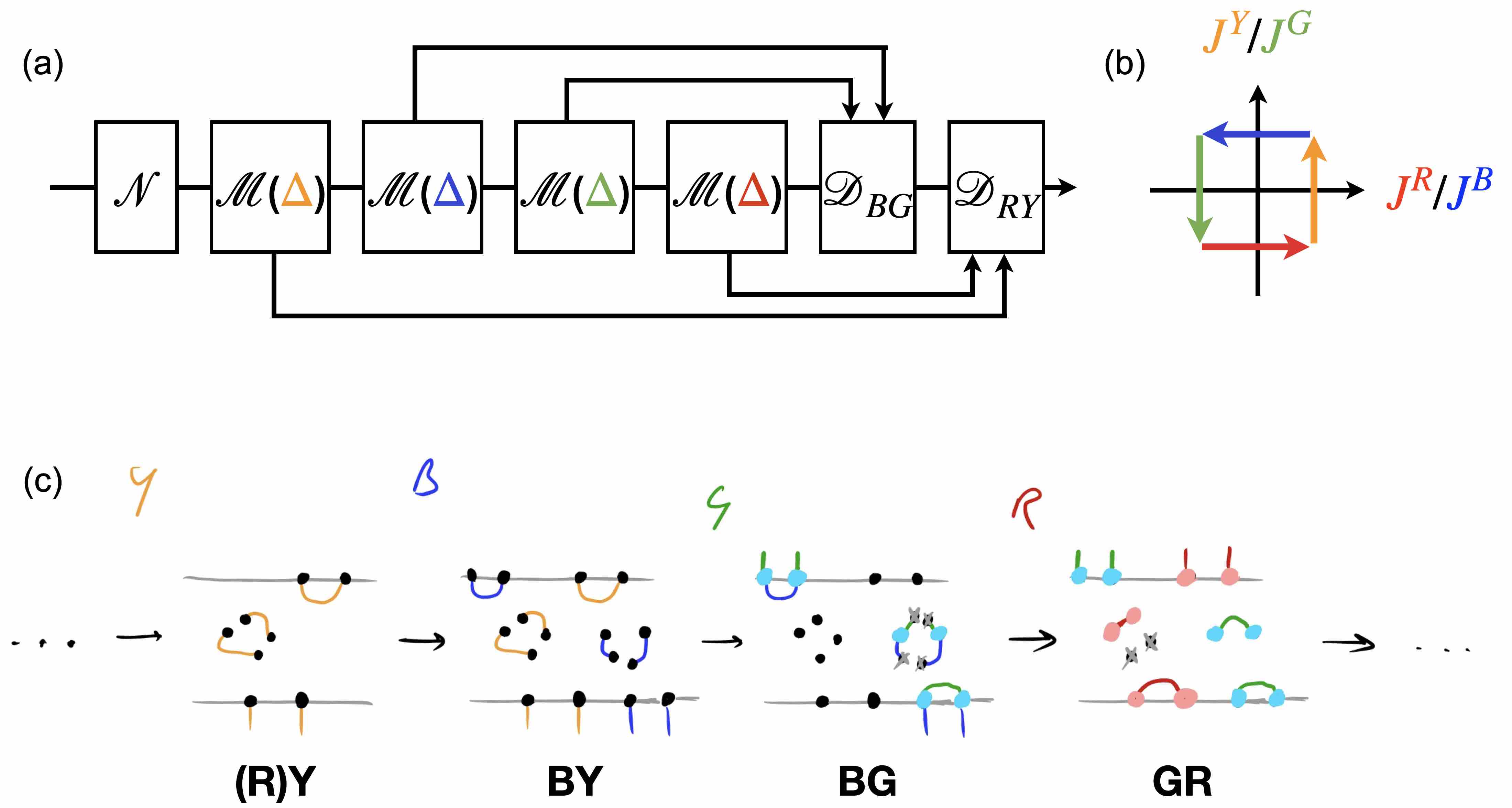}
    \caption{(a) An alternative measurement sequence to Fig.~\ref{fig:ssec}.
    The decoding is performed at the end of the round, after all four types of measurements are completed.
    Since there is only $O(1)$ time between the measurements and the decoding, SSEC is still expected, despite the difference in this protocol from what is preferred in practice, where the decoding is performed as soon as the necessary syndromes are available.
    (b) This sequence induces a different sequence of transitions in the phase diagram, namely $    \ldots
    \rightarrow RY
    \rightarrow BY
    \rightarrow BG
    \rightarrow GR
    \rightarrow RY
    \rightarrow
    \ldots$
    (c) Interestingly, by the time the G checks are measured, the syndrome of the Y checks are randomized.
    However, since the G checks do not generate or move point charges, it suffices to keep track of the endpoints of Y checks, and wait until R checks are measured before performing the matching decoder, $\mathcal{D}_{RY}$.
    In particular, we keep a point charge (and identify it as a stabilizer error) only if it appears at the endpoint of both R and Y fluxes, whereas those appearing at the endpoint of only one flux are identified as measurement errors and neglected thereafter (represented by crossed-out dots in the figure).
    }
    \label{fig:ssec_alt1}
\end{figure*}

\begin{figure*}
    \includegraphics[width=0.8\textwidth]{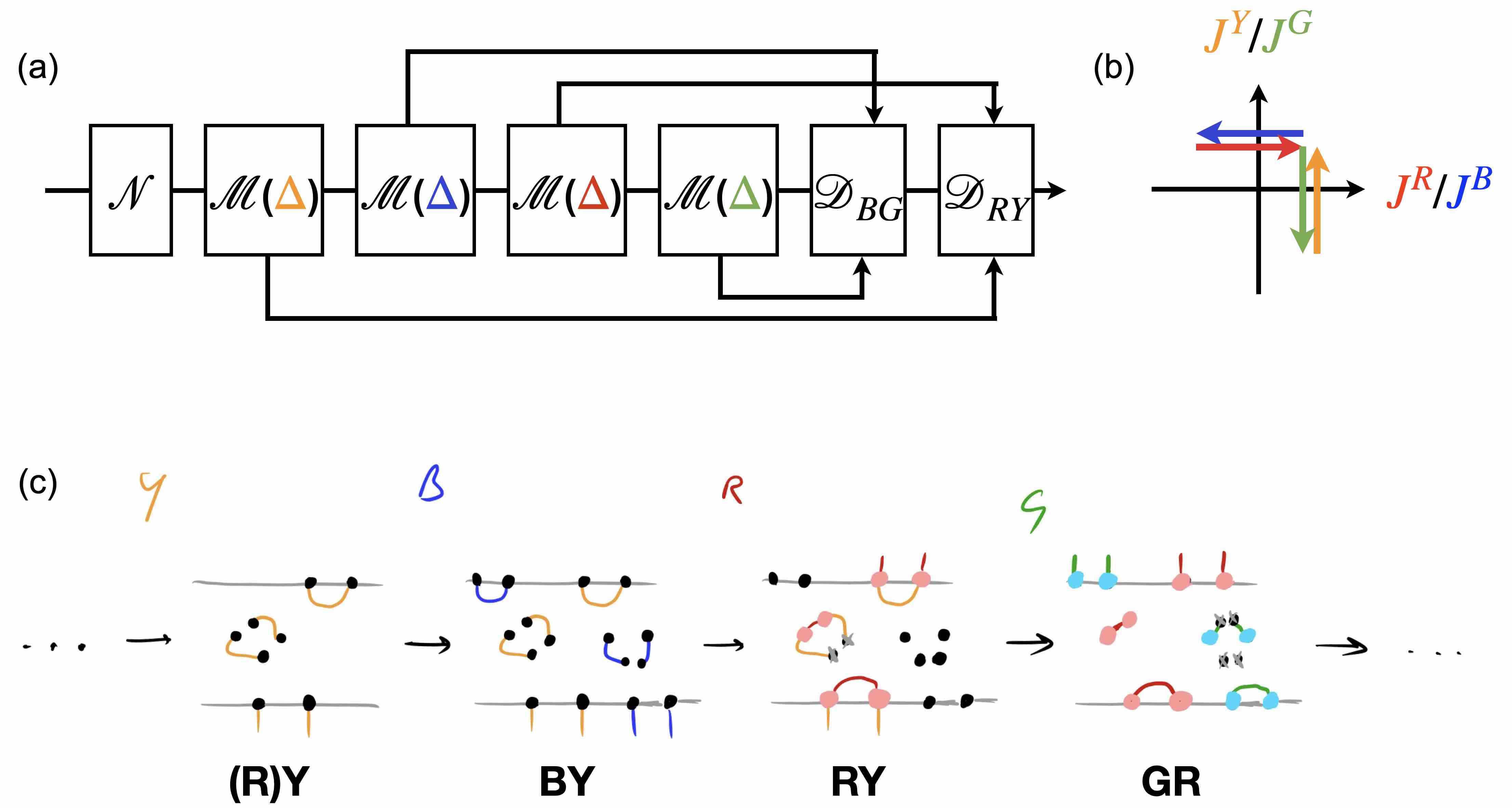}
    \caption{(a) Yet another alternative measurement sequence to Fig.~\ref{fig:ssec}.
    (b) The induced sequence now only visits three phases, namely RY, BY, and GR.
    (c) Similarly to Fig.~\ref{fig:ssec_alt1}, R measurements randomizes B fluxes, but
    the decoding works if endpoints of B fluxes are kept track of and when the matching decoder, $\mathcal{D}_{BG}$ is performed after G measurements.}
    \label{fig:ssec_alt2}
\end{figure*}

Our above results show that single-shot error correction (abbreviated SSEC below) does not imply the existence of any novel or thermally robust bulk-phases in the associated Hamiltonian models. However, the codes and the Hamiltonian models share a few features. They have the same Hilbert spaces and descriptions in terms of kinetically constrained flux loops. They moreover involve the same Pauli check operators and, at least at some points in the evolution of the code,  share the same stabilizer and logical operators. These features can be harnessed in conjunction with a decoding algorithm to produce the SSEC properties of the 3D subsystem toric code. In the context of the Hamiltonian model, they do imply some interesting characteristics in the phase diagram (e.g., the kinetic constraint is responsible for the total bulk confinement in the off-diagonal phase), but they are insufficient to imply thermal stability.

In this section we detail how, and at what stages of the algorithm, the 3D subsystem toric code shares the same stabilizer and logical operators as the Hamiltonian models. This task, it turns out, is akin to associating the 3D subsystem toric code (and its variants) as a path through the zero temperature STC phase diagram. This is somewhat similar to the description of 2D Floquet codes\cite{hastings2021FloquetCode, aasen2022adiabatic, davydova2022floquet, brown-2022-anyon-condensation} in terms of sequence of phase transitions~\cite{brown-2022-anyon-condensation}.

To begin, we recall the ``standard sequence'' for SSEC proposed by Kubica and Vasmer~\cite{KubicaVasmer}.
First, all the $X$ checks (red and yellow) are measured. Next all $Z$ checks (blue and green) are measured.
We can represent this sequence pictorally in Fig.~\ref{fig:ssec}(a), and algebraically as
\begin{align}
\label{eq:ssec_seq_standard}
    \ldots
    \rightarrow RY
    \rightarrow BG
    \rightarrow RY
    \rightarrow BG
    \rightarrow
    \ldots
\end{align}
This sequence can be thought of as jumping between the two diagonal phases (BG and RY), see Fig.~\ref{fig:ssec}(b). For example, immediately following the RY checks, the resulting state will be an eigenstate of Eq.~\eqref{eq:Hamiltonian_J_K} with just the BG couplings switched off. The resulting instantaneous state is related to the ground state of the $J^{R,Y}\neq0,J^{R,B}=0$ phase by a low-depth circuit, thus we represent it with a corresponding point in the  phase diagram. Similarly, after the BG checks are performed, the resulting state will  be an eigenstate of Eq.~\eqref{eq:Hamiltonian_J_K} when the RY couplings are switched off; we can represent the instantaneous state of the system by a point in the $J^{R,B}\neq0,J^{R,Y}=0$ phase.

We now discuss how the above procedure leads to SSEC.
After the RY checks are measured, the R and Y fluxes together must form closed loops, and locations where a loop changes color are identified with point charges (i.e. stabilizer violations induced by phase-flip $Z$ errors).
With stochastic noise and when the error rate is low, the point charges will be closely bound in pairs, but the fluxes will typically fluctuate wildly and can extend throughout the entire system (i.e. closed R and Y loops ``condense'').
An example is shown in Fig.~\ref{fig:ssec}(d).
Imperfect check measurements lead to missing segments of the closed loops, as well as short, isolated flux lines, as we illustrate in Fig.~\ref{fig:ssec}(c). 
The decoder for R and Y fluxes (denoted $\mathcal{D}_{RY}$ in Fig.~\ref{fig:ssec}) can account for imperfect measurements using a three-step algorithm, namely 1) close all loops by pairing up endpoints of missing segments,\footnote{In this step the pairing algorithm does not distinguish between endpoints of R fluxes from those of Y fluxes. 
That is, any endpoint is allowed to be paired with any other.
In the case when an R flux and a Y flux terminates on the same cube stabilizer, we treat this as a double degeneracy, and it is always acceptable to pair them up at zero cost.}
2) identify stabilizer errors as where the loop changes color, and 3) pair up stabilizer errors.
The pairing subroutine, for both steps 1 and 3, can be chosen to be the minimum-weight perfect matching algorithm~\cite{KubicaVasmer}.
The physical correction operation is applied based on the choice of path connecting two error syndromes in step 3.
One can see that if we only measure R checks but not the Y checks, it is impossible to distinguish stabilizer errors from measurement errors, as they both appear as short, missing segments of a long, winding loop.

The standard sequence Eq.~\eqref{eq:ssec_seq_standard} traversing between the BG and RY phases illustrates the difficulty in attributing SSEC to a particular phase of the STC Hamiltonian.
It is natural to ask whether different paths through the phase diagram also lead to SSEC.

One possible such sequence is shown in Fig.~\ref{fig:ssec_alt1}(a), where in each round there are four steps, and within each step we measure all checks with one of the four colors.
The colors are ordered periodically as follows,
\begin{align}
    \label{eq:ssec_seq_alt1}
    \ldots
    \rightarrow Y
    \rightarrow B
    \rightarrow G
    \rightarrow R
    \rightarrow Y
    \rightarrow
    \ldots
\end{align}
This sequence induces the following sequence of phase transitions,
\begin{align}
    \ldots
    \rightarrow RY
    \rightarrow BY
    \rightarrow BG
    \rightarrow GR
    \rightarrow RY
    \rightarrow
    \ldots
\end{align}
see Fig.~\ref{fig:ssec_alt1}(b,c).
Within one round, measurements of $G$ would partially randomize the $Y$ syndrome, and similarly measurements of $R$ would partially randomize the $B$ syndrome.
However, since the checks commute with all the stabilizers, they do not introduce new stabilizer errors and therefore do not affect the point charges where the fluxes end.
Thus, we can still decode by comparing endpoints of missing segments of closed loop, even if these endpoints are read out at different times.
An example is shown in Fig.~\ref{fig:ssec_alt1}(c), where we collect the endpoints of the fluxes at each step, and compare R endpoints with Y endpoints (and similarly for B and G endpoints).
Those appearing at the end of both R and Y fluxes are identified as stabilizer errors, and the others are identified as readout errors and thereafter neglected (represented by a crossed-out dot in Fig.~\ref{fig:ssec_alt1}(c)).
This holds despite the fact that by the time R checks are measured (step 4), the Y syndrome have already been randomized by G check measurements in step 3.

As yet another example, we consider the sequence as shown Fig.~\ref{fig:ssec_alt2},
\begin{align}
    \label{eq:ssec_seq_alt2}
    \ldots
    \rightarrow Y
    \rightarrow B
    \rightarrow R
    \rightarrow G
    \rightarrow Y
    \rightarrow
    \ldots
\end{align}
It is obtained from Eq.~\eqref{eq:ssec_seq_alt1} by exchanging the order of $R$ and $G$.
It induces the following sequence of phases,
\begin{align}
    \ldots
    \rightarrow RY
    \rightarrow BY
    \rightarrow RY
    \rightarrow GR
    \rightarrow RY
    \rightarrow
    \ldots
\end{align}
The BG phase is never visited, but as we can see from Fig.~\ref{fig:ssec_alt2} it is also good for SSEC. 

To summarize our discussion here, what is important for SSEC is that all four colors of checks are measured within a single round, but the ordering is arbitrary.
More generally, when performing decoding of both BG and RY, it is sufficient that the individual colors have been measured within a constant number of the previous rounds. It is also acceptable if the decoding is not performed immediately after all necessary colors have been measured, but rather performed at the end of the round, as we illustrated in Figs.~\ref{fig:ssec_alt1}, \ref{fig:ssec_alt2}.
We can also choose different orderings of the four colors in different rounds.
We have emphasized the difficulty in relating SSEC to any particular phase of the Hamiltonian, as for any given phase there exists a measurement sequence that never visits that phase, in a manner similar to Fig.~\ref{fig:ssec_alt2}.
We can phrase the process of SSEC as a sequence of transitions across the phase diagram, but to understand
the success of SSEC (and the existence of a threshold) a proof~\cite{bombin2015singleshot, KubicaVasmer}, or a classical stat mech model along the lines of Ref.~\cite{DKLP2001topologicalQmemory, bombin2015singleshot, ChubbFlammia2021} (if it can be written down), is most likely still necessary.

\section{Discussions \label{sec:discussion}}

We showed that the 3D subsystem toric code does not lead to thermally stable quantum memories when interpreted as a family of Hamiltonian models. This is a counterexample to a conjecture~\cite{bombin2015gaugecolorcode, brown2016selfcorrectingRMP, preskill2017qec} that single-shot error correcting codes may correspond to thermally stable topological order in a Hamiltonian model.
Moreover, in its zero temperature phase diagram, we find no ``exotic'' phases in the Hamiltonian models that cannot be otherwise realized in previously known models, e.g. subspace toric codes.

Nevertheless, at zero temperature, the Hamiltonians lead to several somewhat unusual presentations of known topologically ordered phases. One type of  bulk phase (so-called diagonal, following Fig.~\ref{fig:offdiagonal_phase_diagram}) consists of two bulk 3D $\mb{Z}_2$ gauge theories with coupled Gauss laws described above.
In the other type of phase (off-diagonal), a specific (albeit natural) open boundary condition leads to 2D toric code order which is confined to one of the boundaries.
Later in Appendix~\ref{sec:bc_wrong}, we describe alternative boundary conditions which lead to no logical qubits.
Thus there is no sense in which the boundary topological phase is forced upon us by the bulk phase --- indeed, in the off-diagonal phases (BY and GR) of Fig.~\ref{fig:offdiagonal_phase_diagram}, the bulk and boundary degrees of freedom can be completely disentangled by a finite-depth unitary circuit.

All of the Hamiltonians we consider have membrane symmetries. These form a non-trivial algebra for the correct boundary conditions described in Sec.~\ref{sec:bare_logicals}, which in turn implies the existence of a ground state degeneracy throughout the phase diagram.
In the off-diagonal phase, said membrane operators can be reduced to line-like operators which act at the boundary, using the check conditions enforced by the ground state.

\subsection{Other notable subsystem codes}

Although we have been focusing on one particular model, it seems possible that much of our analysis here might carry over to the gauge color code and yield the conclusion of the absence of a self-correcting phase, at least for the simplest Hamiltonian~\cite{burton2018spectra}.
The two models share the same essential ingredients.
Our ``effective Hamiltonian'' approach is similar to the ``ungauging'' approach by Kubica and Yoshida~\cite{kubica2018ungauging} for the gauge color code, for which they found six copies of decoupled lattice gauge theory on a cubic lattice. 
The kinetic constraint for the STC that couples the two LGTs also has an immediate analog in the GCC, namely the ``color conservation'' condition.

The gauge color code and the subsystem toric code are special types of subsystem codes in that the stabilizers are local, and that naturally leads to a local gauge invariance.
Other subsystem codes, such as Bacon-Shor codes~\cite{bacon2006compass}, has nonlocal stabilizers and the physics should be entirely different~\cite{ioffe2005compass, becca2005compassmodel, nussinov2015compassrmp}.
Whether they can be thermally stable memories remains an open question~\cite{brown2016selfcorrectingRMP}.

On a technical note, CSS Hamiltonians do not have a sign problem in the computational basis, and can be accessed by quantum Monte Carlo methods.
Direct numerical results of CSS subsystem code Hamiltonians (for the 3D STC and for others) would be an interesting contribution to the study of these systems.

\section*{Acknowledgements}

We acknowledge helpful conversations or correspondences with Ehud Altman, Leon Balents, Fiona Burnell, Margarita Davydova, Arpit Dua, Matthew Fisher, Aram Harrow, Ikuo Ichinose, Wenjie Ji, Aleksander Kubica, Tsung-Cheng Lu, Rahul Nandkishore, John Preskill, Tibor Rakovszky, Brian Skinner, Michael Vasmer, and Dominic Williamson.
YL is supported by an IBM research internship, and in part by the Gordon and Betty Moore Foundation’s EPiQS Initiative through Grant GBMF8686, and in part by the Stanford Q-FARM Bloch Postdoctoral Fellowship in Quantum Science and Engineering.  CWvK is supported
by a UKRI Future Leaders Fellowship MR/T040947/1. GZ and TJO are  supported by the U.S. Department of Energy, Office of Science, National Quantum Information Science Research Centers, Co-design Center for Quantum Advantage (C2QA) under contract number DE-SC0012704. YL and CWvK acknowledge the hospitality of the Kavli Institute for Theoretical Physics at the
University of California, Santa Barbara (supported by the National Science Foundation under Grant No. NSF PHY-1748958).

\emph{Note added}. --- 
We would like to bring the reader’s attention to a related independent work by Bridgeman,  Kubica, and Vasmer~\cite{bridgeman2023}, which generalizes the subsystem toric code construction and appear in the same arXiv posting.

\bibliography{refs}

\appendix

\section{Classical Monte Carlo study \label{sec:classical_MC}}
\begin{figure}[b]
    \centering
    \includegraphics[width=.45\textwidth]{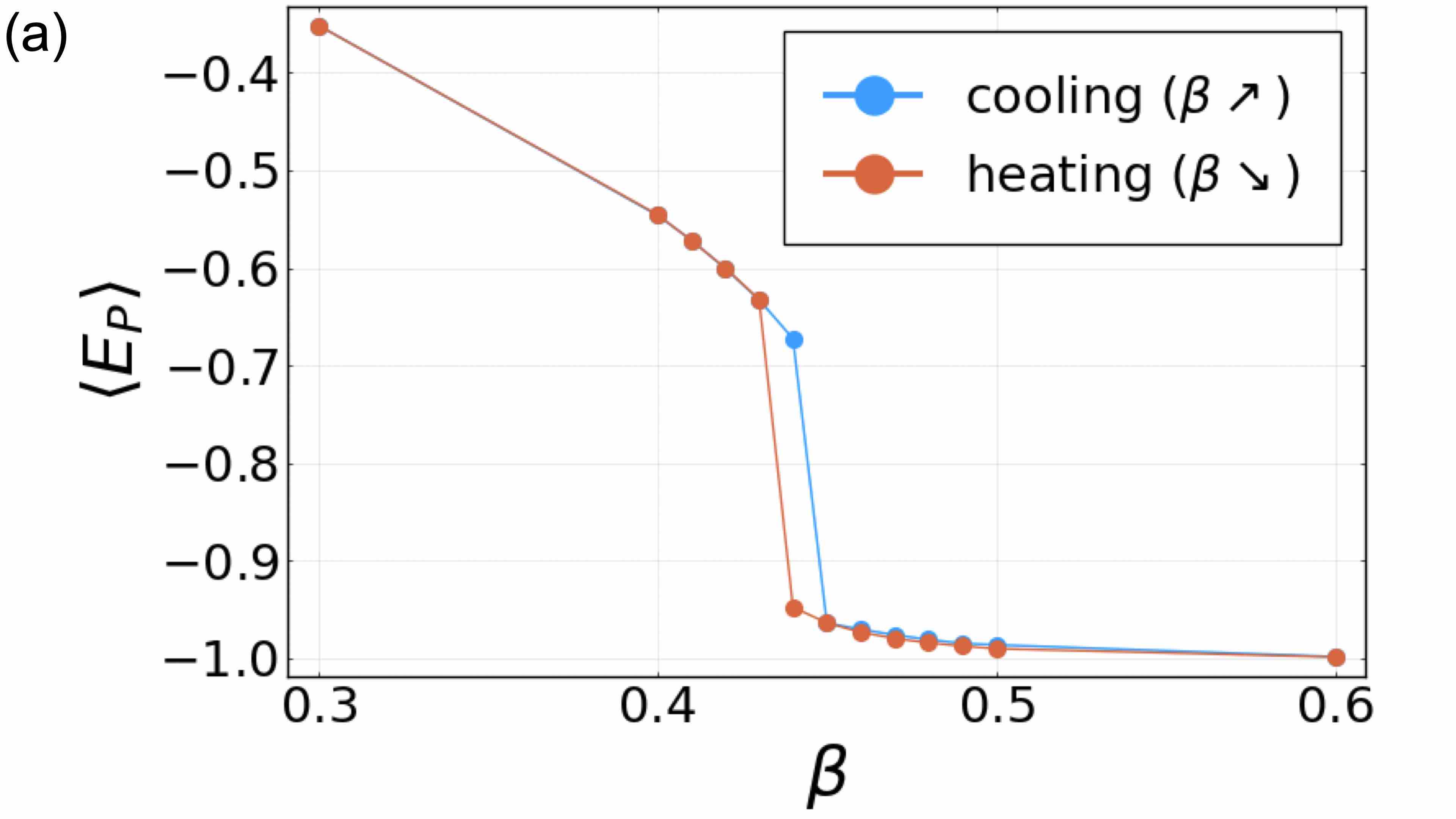}
    \includegraphics[width=.45\textwidth]{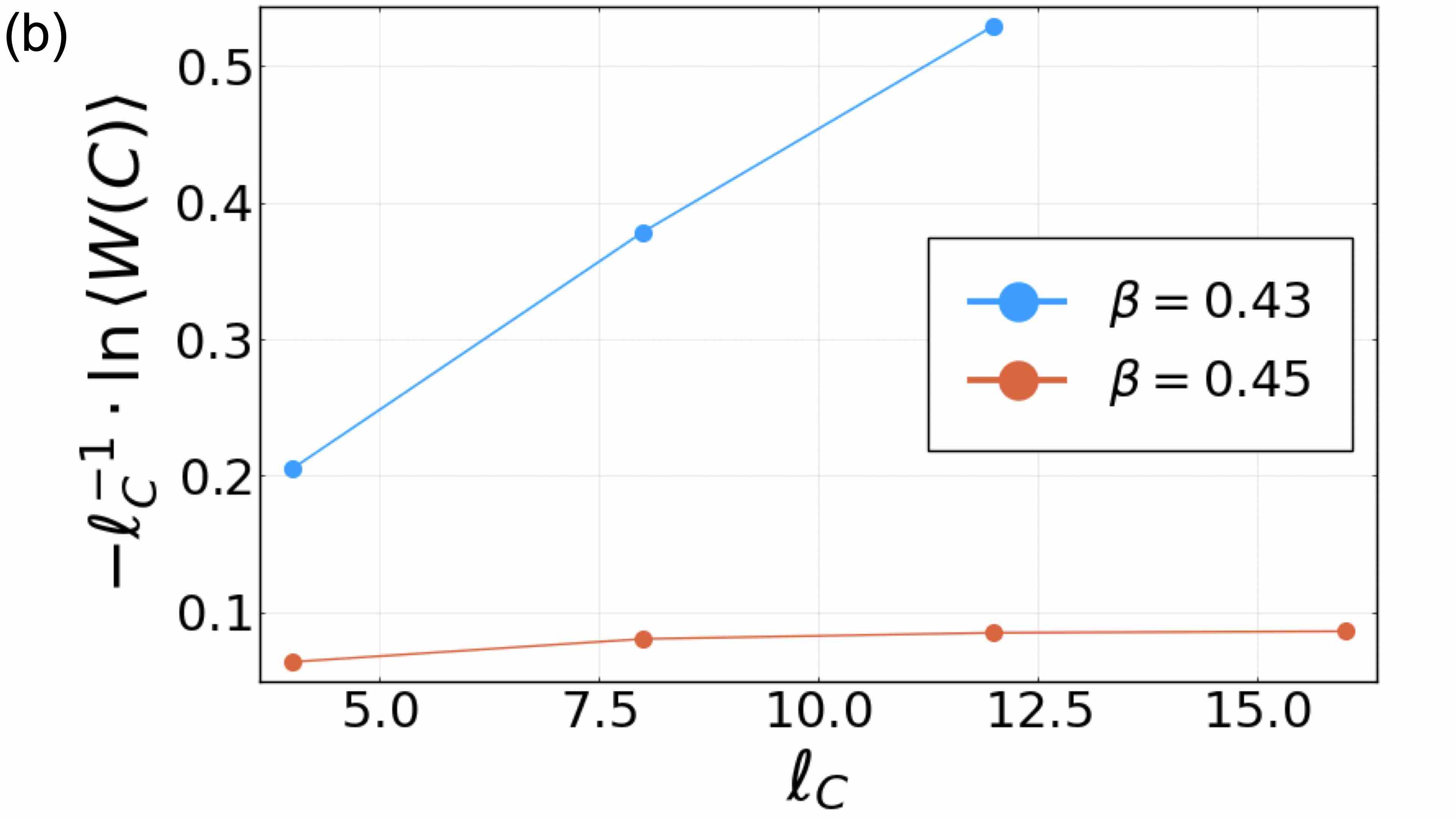}
    \caption{Classical Monte-Carlo data for the classical gauge theory in Eq.~\eqref{eq:classical_gauge_theory}.
    (a) The average energy of the space-like hexagon term, $\avg{\prod_{\avg{ij} \in \hexagon} z_{ij}(\tau)}$.
    The discontinuity and the hysteresis loop are both signatures of a first-order transition~\cite{Creutz1979Ising, Creutz1979ZN, Creutz1983PhysRev, ichinose2005fourdimZ2}.
    (b) The Wilson loop operator is found to exhibit area law and perimeter law scaling, respectively, when $\beta < \beta^\ast$  and $\beta > \beta^\ast$, see Eq.~\eqref{eq:wilson_loop}.
    Here we choose the loop $C$ to have four sides, each with $\ell_C/4$ bonds.
    The area bounded by $C$ is proportional to $\ell_C^2$.
    }
    \label{fig:MonteCarlo}
\end{figure}
In this appendix, we perform a Monte Carlo study of the effective Hamiltonian $H^{\rm LGT}$, see Eq.~\eqref{eq:H_LGT}, to produce the phase diagrams in Fig.~\ref{fig:ZN_phase_diagram}(a) and Fig.~\ref{fig:offdiagonal_phase_diagram}.
The partition function $\mc{Z} = \Tr e^{-\beta H^{\rm LGT}}$ with $H^{\rm LGT}$ from Eq.~\eqref{eq:H_LGT} can be related to that of a classical $\mb{Z}_2$ gauge theory~\cite{fradkinsusskind1978, kogut1979rmp} on a four-dimensional lattice, where the fourth dimension comes from the inverse temperature $\beta = 1/T$.
We treat the partition function as a path integral in imaginary time, and consider discretizing the imaginary time into infinitesimal segments of length $\delta \tau$.
The factors involved in the path integral take the form
\begin{align}
    & \bra{z(\tau_{t+1})} e^{-\delta \tau H^{\rm dual}} \ket{z(\tau_t)} \nn
    \approx&\ \bra{z(\tau_{t+1})}
    e^{\delta \tau (J_X/2) \sum_{\hexagon} \prod_{\avg{ij} \in \hexagon}\tau^z_{ij}}
    \ket{z(\tau_{t+1})} 
    \nn &\quad 
    \cdot \bra{z(\tau_{t+1})} e^{\delta \tau (J_Z/2) \sum_{\avg{ij}} \tau_{ij}^x} \ket{z(\tau_t)} \nn
    \propto &\ e^{\beta_s \sum_{\hexagon} \prod_{\avg{ij} \in \hexagon} z_{ij}(\tau_{t+1})}
    e^{\beta_\tau \sum_{\avg{ij}} z_{ij}(\tau_{t+1}) z_{ij}(\tau_{t})}.
\end{align}
Here $\beta_s = J_X \delta \tau / 2$, and $e^{-2\beta_\tau} = \tanh(J_Z \delta \tau/2)$.
The partition function thus becomes that of a classical statistical mechanics model, namely
\begin{align}
\label{eq:classical_gauge_theory_temporal_gauge}
    \mc{Z}_{\rm cl} \propto& \Tr_{\{z_{ij}(\tau_t)\} } e^{-E/T}, \nn
    E/T =& - \beta_s \sum_{t, \hexagon} \prod_{\avg{ij} \in \hexagon} z_{ij}(\tau_{t})
    - \beta_\tau \sum_{t, \avg{ij}} z_{ij}(\tau_{t+1}) z_{ij}(\tau_{t}).
\end{align}
This partition function is equivalent to the more familiar form of a lattice gauge theory,
\begin{align}
\label{eq:classical_gauge_theory}
    \mc{Z}_{\rm LGT} \propto& \Tr_{\{z_{ij}(\tau_t), z_i(\tau_t, \tau_{t+1})\} } e^{-E/T}, \nn
    E/T =& - \beta_s \sum_{t, \hexagon} \prod_{\avg{ij} \in \hexagon} z_{ij}(\tau_{t}) \nn
    &\quad
    - \beta_\tau \sum_{\avg{ij}} z_{ij}(\tau_{t+1}) z_{ij}(\tau_{t})
    z_{i}(\tau_{t+1}, \tau_{t}) z_{j}(\tau_{t+1}, \tau_{t}).
\end{align}
The partition function $\mc{Z}_{\rm LGT}$ explicitly involves all plaquttes of the 4D lattice.
It is invariant under local gauge transformations that flips the signs of all the variables incident to a given vertex.
The partition function $\mc{Z}_{\rm cl}$ is obtained from $\mc{Z}_{\rm LGT}$ after gauge fixing all temporal links $z_{i}(\tau_{t+1}, \tau_{t})=1$; this can always be achieved with the said gauge transformation.

We can thus study the phase diagram of $H_{\rm LGT}$ by studying that of $\mathcal{Z}_{\rm LGT}$.
The imaginary time path integral is well approximated if $\beta_\tau \to \infty$ and correspondingly $\beta_s \propto e^{-2 \beta_\tau} \to 0$, but it is often possible to choose a different set of $(\beta_s, \beta_\tau)$ with the same qualitative long-distance behavior~\cite{fradkinsusskind1978}.
For the purpose of accessing the phase transition, we can focus on the isotropic case $\beta_s = \beta_\tau = \beta$, so that $\beta$ can be interpreted as the inverse temperature of the classical model.
In Fig.~\ref{fig:MonteCarlo} we provide evidence to the phase diagram Fig.~\ref{fig:ZN_phase_diagram}(a) with classical Monte Carlo data from $\mathcal{Z}_{\rm LGT}$.
We find that the expectation value of an elementary plaquette energy term $E_P = \avg{\prod_{\avg{ij} \in \hexagon_P} z_{ij}(\tau)}$ has a discontinuity near $\beta^\ast \approx 0.44$,
and exhibits a hysteresis loop when the system is cooled or heated across the $\beta^\ast$.
Both indicates a first-order phase transition.
Furthermore, we find that the Wilson loop operator $\avg{W(C)} = \avg{\prod_{b \in C} z_{b}}$ defined for an arbitrary closed loop $C$ in the Euclidean lattice obeys area and perimeter law in the two phases,
\begin{align}
\label{eq:wilson_loop}
    \avg{W(C)} \sim \begin{cases}
        e^{- c(\beta) \cdot \mathrm{Area}(C)}, \quad &\beta < \beta^\ast \\
        e^{- c(\beta) \cdot \mathrm{Perimeter}(C)}, \quad &\beta > \beta^\ast
    \end{cases}.
\end{align}
The area law scaling of $\avg{W(C)}$ is usually used as a diagnostic of confinement.
Thus, other than the fact that $H^{\rm LGT}$ is defined on a diamond lattice (rather than a cubic lattice), our findings here are completely analogous to the cubic lattice $\mb{Z}_2$ gauge theory, see Ref.~\cite{Creutz1979Ising}.

Going back to $H_{\rm STC}$ in Eq.~\eqref{eq:Hamiltonian_J_K}, which is a sum of $H_{BR}$ and $H_{GY}$.
Each of the two can be mapped to $H^{\rm LGT}$ under duality transformations, and the two are decoupled at zero temperature, when all point charges are gapped out.
Thus, the phase diagram of $H_{\rm STC}$ takes a simple product form, as we depicted in Fig.~\ref{fig:offdiagonal_phase_diagram}.
The phase boundaries in Fig.~\ref{fig:offdiagonal_phase_diagram} correspond to a transition in one of the two lattice gauge theories.

\section{Summary of bulk and boundary excitations \label{sec:summary_excitations}}

\begin{figure*}[t]
    \centering
    \includegraphics[width=1.00\textwidth]{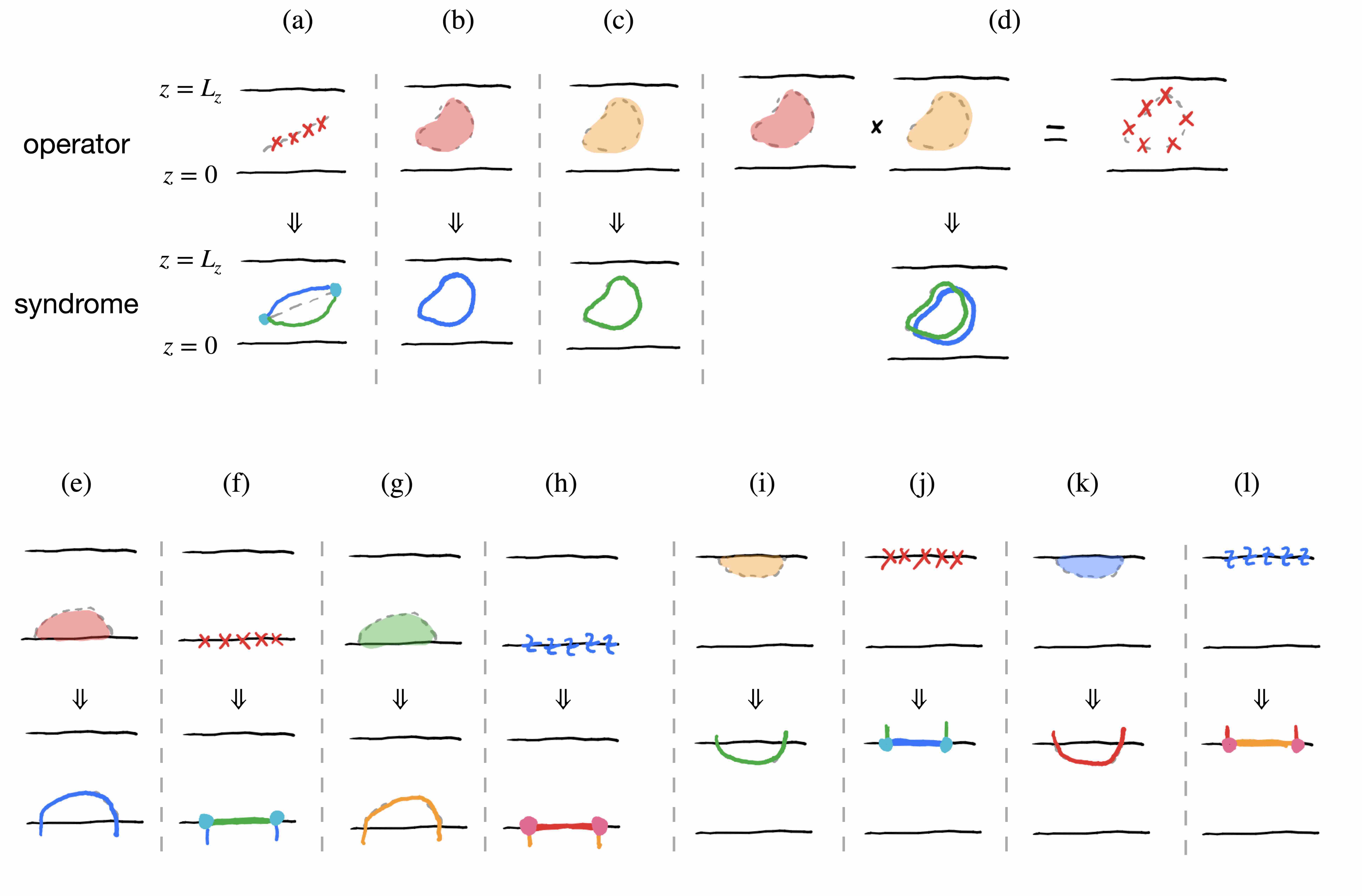}
    \caption{Excitations of the model Eq.~\eqref{eq:Hamiltonian_J_K} in the bulk (a-d) and near the boundary (e-l).
    For each of these, we illustrate both the creation operator and the ``syndrome'' it gives rise to, i.e. the checks that anticommute with the creation operator.
    (a) A string of Pauli $X$ operators creates a pair of $e$ charges (cyan dots) connected by a blue flux string and a green flux string.
    This is a generalization of the microscopic excitation illustrated in Fig.~\ref{fig:TC_excitations}(b).
    There is also the excitation of two $m$ charges as created by a string of Pauli $Z$s, but is not shown here.
    (b,c) A closed surface of red or yellow checks creates a blue or green flux loop at its boundary.
    This is a generalization of the microscopic excitation illustrated in Fig.~\ref{fig:face_to_bond}(c).
    (d) When multiplying the creation operators in (b) and in (c) on the same surface, we have a blue and a green flux loop closely track each other.
    This can be thought of as the ``trace'' of a pair of $e$ charges traversing a closed loop, after which they annihilate.
    This means that the product creation operator is trivial in the interior of the surface, and only has nontrivial content on its boundary, namely a closed loop of Pauli $X$s that create and move the $e$ charges.
    (e-h) Excited states near the lower boundary.
    Among these, (e, f) are generalizations of the microscopic excitations illustrated in Fig.~\ref{fig:boundary_check_to_flux}(c,d).
    (g) is obtained from (e) by replacing the $X$-type red checks with $Z$-type green checks, and (h) from (f) by replacing the string of Pauli $X$s by a string of Pauli $Z$s.
    (i-l) are excitations near the upper boundary, that can be obtained in a way similar to (e-h), namely by taking into account the boundary conditions in Fig.~\ref{fig:boundary_details}. 
    }
    \label{fig:summary_excitations}
\end{figure*}

\begin{figure*}
    \centering
    \includegraphics[width=0.90\textwidth]{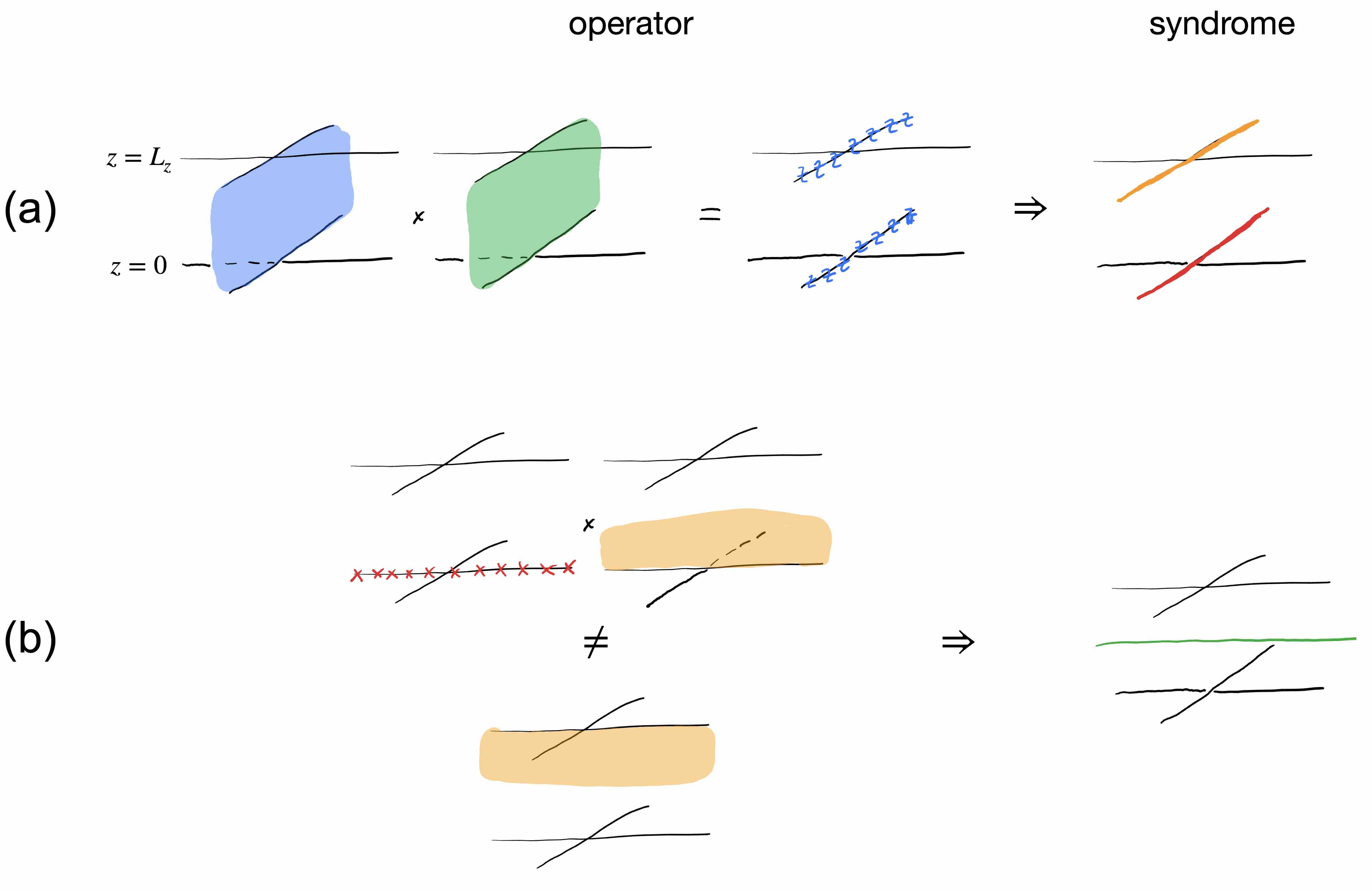}
    \caption{(a) Product of blue checks and green checks on the same surface (streching between the two boundaries) is equivalent two strings of Pauli $Z$s at the intersection of the surface with the upper and lower boundaries, compare Fig.~\ref{fig:summary_excitations}(d,g,k).
    This operator creates a yellow flux string on the upper boundary, and a red flux on the lower boundary, compare Fig.~\ref{fig:summary_excitations}(h,l).
    (b) Composing a string of Pauli $X$s on the lower boundary and a surface of yellow checks results in a single green flux in the bulk, but no other syndromes, compare Fig.~\ref{fig:summary_excitations}(f,c).
    The same syndrome can be created by a different operator, namely a yellow surface from the upper boundary, compare Fig.~\ref{fig:summary_excitations}(i).
    }
    \label{fig:curt}
\end{figure*}

In Fig.~\ref{fig:summary_excitations}, we summarize different types of excitations allowed by the kinetics that can occur at long length scales in at least one phase of the model (with boundary conditions as in Sec.~\ref{sec:boundary_surface_code}), without referring to their energies.
For each of them, we list both the ``creation operator'' and the ``syndrome'' it gives rise to, i.e. the checks and stabilizers that anticommute with the creation operator.
Most of these have been introduced previously in Figs.~(\ref{fig:TC_excitations}, \ref{fig:face_to_bond}, \ref{fig:boundary_check_to_flux}).
Because of the pictures we developed there, we can discuss them without reference to lattice details.

We note that, in any given phase of the STC Hamiltonian, some of these excitations will cost an energy extensive in their sizes, and are in no sense elementary.
So, while all are allowed by kinetics, in discussing any given phase only a subset will be relevant.
This summary is thus most useful as a list of operator identities.

\subsection{A boundary operator detecting bulk flux}

We discuss an apparent puzzle brought by two particular types of excitations, as we describe in Fig.~\ref{fig:curt}.
We denote by $P_{BG}(S)$ the operator in Fig.~\ref{fig:curt}(a), as it is the product of blue checks and green checks on a topologically nontrivial surface $S$ that connects the upper and lower boundaries, thus detecting the parity of the total blue and green flux through $S$.
The operator in Fig.~\ref{fig:curt}(b) is the composition of a string of Pauli $X$s on the lower boundary and a surface of yellow checks, both intersecting with $S$.
We denote this operator $G(\ell)$, where $\ell$ is a line bounding the yellow surface, because the resulting state $G(\ell) \ket{\Omega}$ has a single green flux along $\ell$ through $S$, but there are no other syndromes.
We thus have
\begin{align}
    \bra{\Omega} G(\ell)^\dg \cdot   P_{BG}(S) \cdot G(\ell) \ket{\Omega}  =&\, -1.
\end{align}
On the other hand, from Fig.~\ref{fig:summary_excitations}(d), we notice that $P_{BG}(S)$ acts nontrivially (as Pauli $Z$s) only on $\partial S$, which contains qubits disjoint from $\ell$.
It is thus at least a bit strange that an operator supported only on the boundary can detect a flux in the bulk, on a completely disjoint region.
There is of course no contradiction, since the flux creation operator $G(\ell)$ involves a string of Pauli $X$s on the boundary that is responsible for the anticommutation with $P_{BG}(S)$.
In this model, to create a single flux in the bulk, the operator is necessarily supported nontrivially on the boundary.

As another example, the same syndrome can be created by a \emph{different} operator, denoted $G^\p(\ell)$, which is a surface of yellow checks starting from the upper boundary, see also Fig.~\ref{fig:curt}(b).
The anticommutation with $P_{BG}(S)$ can be checked by noticing that $G^\p(\ell)$ also contains Pauli $Z$s on the upper boundary, see Fig.~\ref{fig:boundary_details}(b).
The product $G^\p(\ell) \cdot G(\ell)$ leaves no syndromes, and acts as a bare logical operator of the code, see Fig.~\ref{fig:bare_logicals}.

\subsection{Dressed logical operators \label{sec:dressed_logicals}}

\begin{figure}
    \centering
    \includegraphics[width=.5\textwidth]{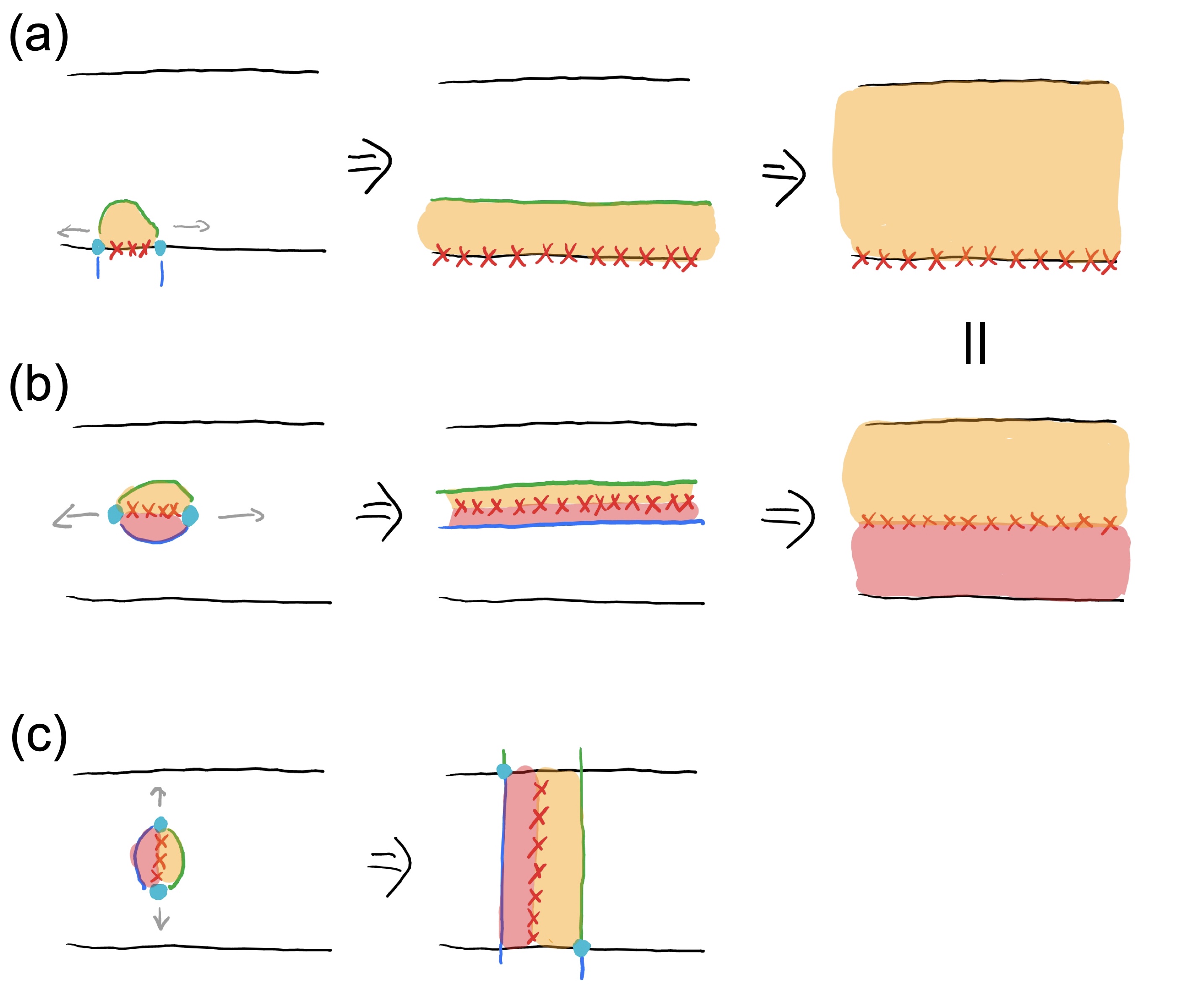}
    \caption{
    (a) A dressed logical operator can be created by a closed loop of Pauli $X$s on the lower boundary. The syndrome it leaves is a green flux string.
    When we multiply this operator by a surface of yellow checks, so as to annihilate the green flux string at the upper boundary, we obtain a bare logical operator, which is identical to the one in Fig.~\ref{fig:bare_logicals}.
    (b) Another dressed logical operator can be created by a closed loop of Pauli $X$s in the bulk.
    The syndrome it leaves is a pair of blue and green flux strings.
    They blue and green fluxes can be similarly annihilated by the boundaries, in which case we again obtain a bare logical operator from the dressed logical operator by multiplying a surface of red or yellow checks.
    The bare logical operator obtained this way is identical to the one in (a), as we can see using the operator identity in Fig.~\ref{fig:summary_excitations}(d).
    (c) Extend the Pauli $X$ string in the $z$ direction fails to produce a dressed logical operator, due to a pair of separated $e$ charges that cannot be annihilated.
    }
    \label{fig:dressed_logicals}
\end{figure}

Besides the bare logical operators discussed in Sec.~\ref{sec:bare_logicals}, ``dressed'' logical operators are also important in the discussion of subsystem codes.
The ``dressed'' logical group is defined as
\begin{align}
    \mc{L}^{\rm dressed} = Z(\mc{S}) / \mc{G}.
\end{align}
The dressed logical operators may anticommute with some of the checks (thus they must create excited states with nonzero fluxes), but can have lower weights.
They differ from bare logical operators by checks, and can mimick the actions of bare ones on the logical information.
Due to their low weight, they are closely related to noises that introduce logical errors in an active error correction setting.

In our case the dressed logicals can be chosen to be (i) identical to the logical operators of the surface 2D toric codes, as we illustrate in Fig.~\ref{fig:dressed_logicals}(a); or they can be chosen to be (ii) noncontractible strings in the bulk, see Fig.~\ref{fig:dressed_logicals}(b), much like two of the 3D toric code logicals.
These two types of dressed logicals can also be understood in terms of the fluxes.
\begin{itemize}
\item 
Type (i) dressed logicals can be thought of as the result of a process illustrated in Fig.~\ref{fig:dressed_logicals}(a).
Initially, a Pauli $X$ string operator on the lower surface creates a (green)flux-(cyan)charge-(blue)anyon composite (same as in Fig.~\ref{fig:bare_logicals}(c)).
Extending the Pauli string to a noncontractible loop around lower surface annihilates the charges and the boundary anyons, but leaves out a noncontratible loop of green fluxes in the bulk.
The green flux loop can be moved and absorbed by the upper surface, and we get the same bare logical operator in Fig.~\ref{fig:bare_logicals}(c).
\item
The creation of a type (ii) dressed logical is illustrated in detail in Fig.~\ref{fig:dressed_logicals}(b).
In the bulk, a Pauli $X$ string operator creates two charges connected by a green and a blue flux string, see also Fig.~\ref{fig:TC_excitations}(b).
If we extend the Pauli string to a noncontractible loop in the bulk along the $x$ or $y$ directions,  thus annihilating the charges, we are left with two closed flux loops.
Similarly to Fig.~\ref{fig:dressed_logicals}(a), the green flux can be moved to the upper boundary and absorbed, and the blue flux can be absorbed by the lower boundary.
We are left with a bare logical operator, that is identical to the one in Fig.~\ref{fig:dressed_logicals}(a), using Fig.~\ref{fig:summary_excitations}(d).
\end{itemize}

On the other hand, due to the choice of boundary condition, a string of Pauli $X$s connecting the two boundaries (as in Fig.~\ref{fig:dressed_logicals}(c)) creates two $e$ charges on the two boundaries, which cannot be brought together and annihilated.
It thus fails to commute with all stabilizers, and cannot be in $Z(\mathcal{S})$.

\section{Further discussions on boundary conditions \label{sec:bc_further_discussions}}

\subsection{Reverse engineering the boundary condition}

\begin{figure}[t]
    \centering
    \includegraphics[width=0.30\textwidth]{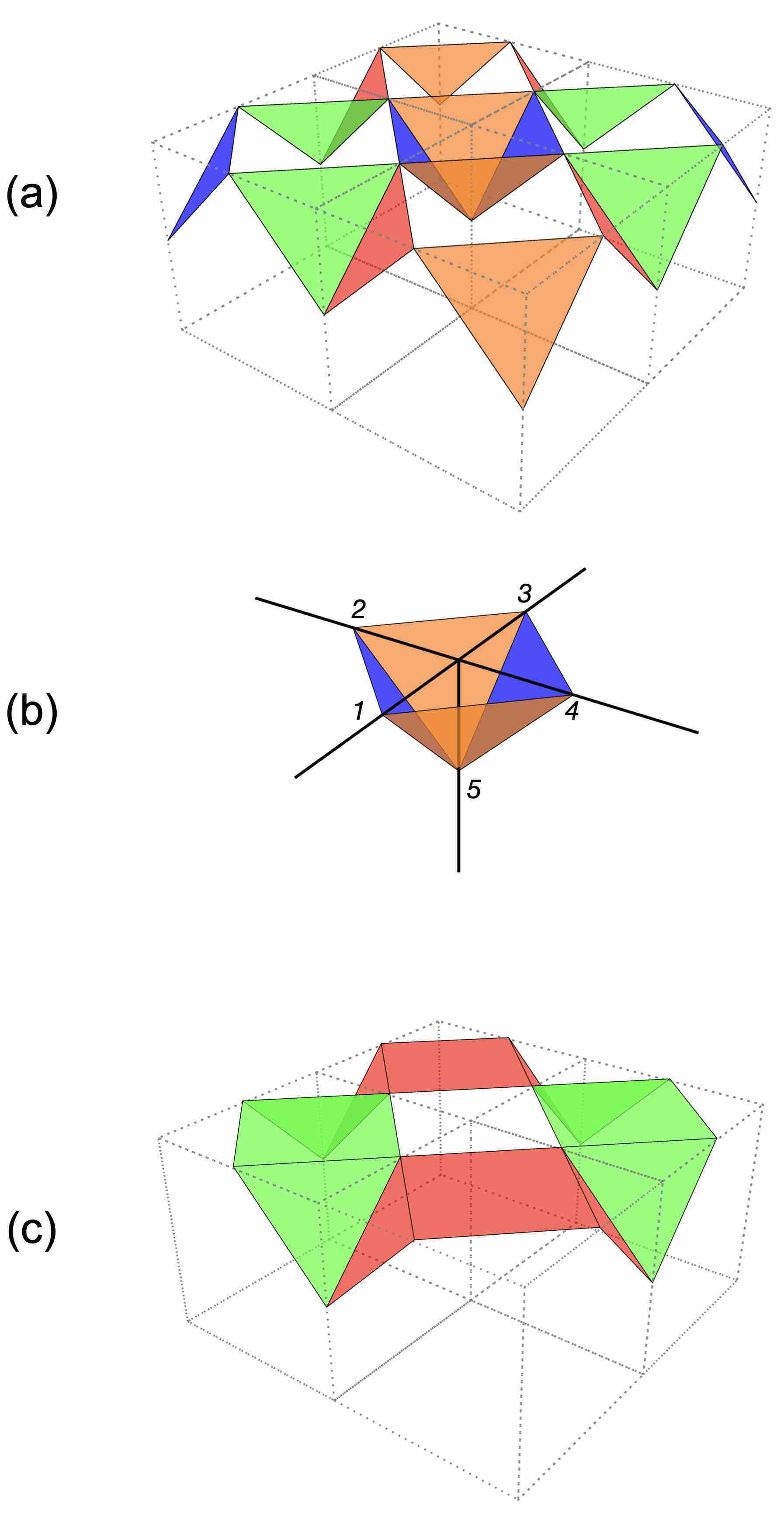}
    \caption{(a) Check operators on the boundary when we cut the lattice open in the $z$ direction at an integer $z$.
    (b) At each ``BY'' type vertex on the boundary, there are two blue and two yellow checks.
    The physical condition that blue and yellow fluxes cannot leave the system imposes the constraints $Z_1 Z_2 Z_3 Z_4 = 1$ and $X_1 X_2 X_3 X_4 = 1$.
    (c) The effective Hamiltonian in the (degenerate) constrained subspace does not have 3-qubit green and red checks.
    Instead, products of two green checks or two red checks, such as those 6-qubit terms shown here, can appear in the effective Hamiltonian.
    }
    \label{fig:bc_reverse_engineer}
\end{figure}

\begin{figure*}[t]
    \centering
    \includegraphics[width=0.8\textwidth]{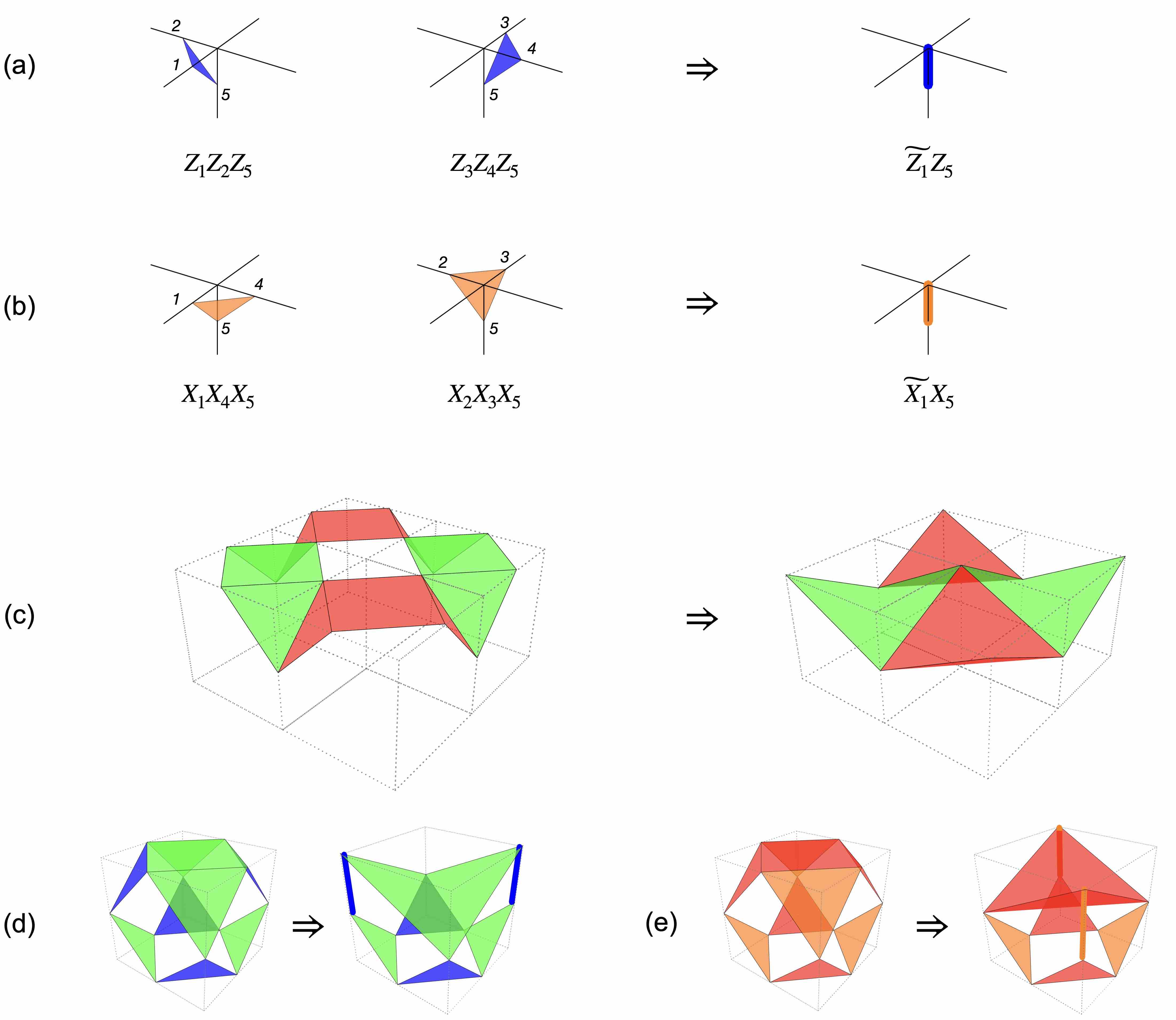}
    \caption{
    Change of variables for the effective Hamiltonian, as described in Eq.~\eqref{eq:disentangling_unitary}.
    (a) The 3-qubit blue checks maps to 2-qubit blue checks, and (b) 3-qubit yellow checks maps to 2-qubit yelow checks.
    In (c), we see that the 6-qubit terms now become 4-qubit terms.
    In (d, e), we collect all terms near the upper boundary.
    We see that the result is identical to Fig.~\ref{fig:boundary_details}(a,b), except that the locations of the qubits are slightly different.}
    \label{fig:bc_reverse_engineer_result}
\end{figure*}

The boundary condition in Fig.~\ref{fig:boundary_details} involves a somewhat \textit{ad hoc} truncation of the lattice, and the equally \textit{ad hoc} introduction of 2 and 4 qubit checks. Here we show that this boundary condition can be reinterpreted more naturally as an interface on the original lattice that is impermeable to two of the flux-types. 

Consider cutting the periodic lattice open in the $z$ direction along two $xy$ planes, at two integer $z$ coordinates.
We focus on the upper boundary (see Fig.~\ref{fig:bc_reverse_engineer}(a)), and try to mimick the boundary condition in Fig.~\ref{fig:boundary_details}(a,b).
In this case, blue and yellow fluxes are forbidden to get out of the plane.
Equivalently, any blue or yellow flux coming into the boundary must be reflected back into the bulk.
On a vertex as shown in Fig.~\ref{fig:bc_reverse_engineer}(b), this is achieved by imposing the following equations as hard constraints:
\begin{align}
    Z_1 Z_2 Z_5 \cdot Z_3 Z_4 Z_5 =&\, Z_1 Z_2 Z_3 Z_4 = \square_Z =  1, \\
    X_1 X_4 Z_5 \cdot X_2 X_3 Z_5 =&\, X_1 X_2 X_3 X_4 = \square_X = 1.
\end{align}
We can add these terms at every vertex of the type in Fig.~\ref{fig:bc_reverse_engineer}(b) (i.e. those with blue and yellow checks) to the Hamiltonian, and associate infinite coupling strengths to them.
For concreteness, let us first fix the notations.
Let
\begin{subequations}
\begin{align}
    H_0 =&\, K_0 \sum_{v \in {\rm upper\ boundary}} [(\square_Z)_v + (\square_X)_v], \\
    H_1 =&\, H_{\rm STC}, 
\end{align}
\end{subequations}
where $H_{\rm STC}$ is the Hamiltonian from Eq.~\eqref{eq:Hamiltonian_J_K}, having coupling strengths $J$ and $K$, where $J, K \ll K_0$.
The notation means that we treat the (mutually commuting) constraints as the unperturbed Hamiltonian, and the STC Hamiltonian as a perturbation.
Let $P$ be the projector onto the degenerate ground space of $H_0$ and $Q = 1-P$.
The effective Hamiltonian follows immediately from degenerate perturbation theory
\begin{align}
    &\ H_{\rm eff} \nn 
    =&\ E_0 + P H_1 P + P H_1 Q \frac{1}{E_0 - H_0}Q H_1 P + O\(\frac{J^3}{K_0^2}, \frac{K^3}{K_0^2}\).
\end{align}
Here $E_0$ is the ground energy of $H_0$. 
All the blue and yellow check terms as well as stabilizer terms in $H_1$ commute with $H_0$, thus survive the projection operator and appear at first order in $H_{\rm eff}$.
On the other hand, green and red checks near the upper boundary (e.g. those shown in Fig.~\ref{fig:bc_reverse_engineer}(a)) anticommute with terms in $H_0$, and are thus eliminated by $P$.
To second order, products of two green checks or two red checks, such as those 6-qubit terms shown in Fig.~\ref{fig:bc_reverse_engineer}(c), commute with $H_0$ and appear in $H_{\rm eff}$, with a strength $O(J^2/K_0)$.
Unlike $\square_Z$ and $\square_X$, the 6-qubit red and green terms do not reflect green and red fluxes, but rather give a gap to green and red fluxes leaving the boundary.

Finally, within the constrained space of $\square_Z = \square_X = 1$, we can treat the constraints as operator identities.
This allows us to perform a change of variables, so that terms in the effective Hamiltonian become more familiar terms in Fig.~\ref{fig:boundary_details}.
Concretely, for each vertex of type in Fig.~\ref{fig:bc_reverse_engineer}(b), one can construct a unitary transformation such that
\begin{subequations}
\label{eq:disentangling_unitary}
\begin{align}
    Z_1 Z_2 \to&\ \widetilde{Z}_1,  \\
    Z_3 Z_4 \to&\ \widetilde{Z}_1 \widetilde{Z}_2, \\
    \square_Z = Z_1 Z_2 Z_3 Z_4 \to&\ \widetilde{Z}_2, \\
    X_1 X_4 \to&\ \widetilde{X}_1,  \\
    X_2 X_3 \to&\ \widetilde{X}_1 \widetilde{X}_3, \\
    \square_X = X_1 X_2 X_3 X_4 \to&\ \widetilde{X}_3.
\end{align}
\end{subequations}
Such a transformation exists as it preserves the algebra of the operators, although it is not completely fixed by these equations.
Within the constrained space of $\square_Z = \square_X = 1$, we can simply treat $Z_1 Z_2 =  Z_3 Z_4 =  \widetilde{Z}_1$ and $X_1 X_4 =  X_2 X_3 =  \widetilde{X}_1$.
The operators also adopt a different form under this new set of variable, as we illustrate in Fig.~\ref{fig:bc_reverse_engineer_result}.
This new representation is identical to Fig.~\ref{fig:boundary_details}.

\subsection{``Wrong'' boundary conditions \label{sec:bc_wrong}}

\begin{figure}[t]
    \centering
\includegraphics[width=.30\textwidth]{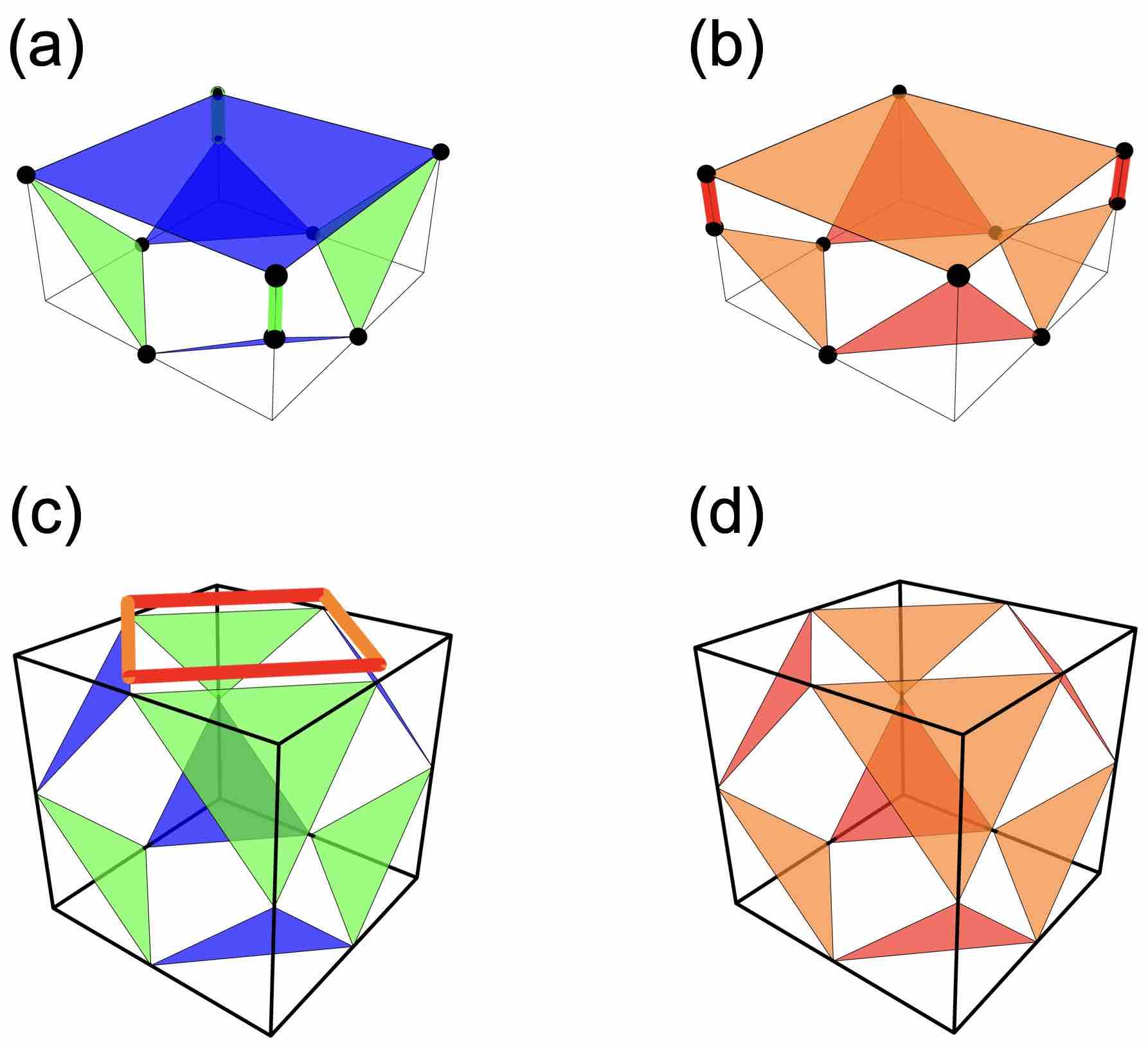}
    \caption{``Wrong'' boundary conditions of the STC that do not lead to logical qubits.
    If we choose the upper boundary conditions as in (a,b), or as in (c,d), the STC Hamiltonian has a unique ground state in the GR phase, and we can no longer construct bare logical operators as Sec.~\ref{sec:bare_logicals}.
    In both cases we keep the lower boundary as before, in Fig.~\ref{fig:boundary_details}(c,d).
    See the text for further discussions.
    }
    \label{fig:boundary_details_wrong}
\end{figure}

Here we discuss two examples of boundary conditions that fail to give the code logical qubits.
In both cases, we only alter the upper boundary conditions and keep the the lower boundary condition unchaged from Fig.~\ref{fig:boundary_details}(c,d).

In the first example, we choose the upper boundary condition
such that the allowed anyons are yellow and blue, just like the lower boundary, see  Fig.~\ref{fig:boundary_details_wrong}(a,b).
In the GR phase, blue and yellow anyons can condense on both boundaries, and the Hamiltonian does not have a degenerate ground space.

In the second example, we choose the upper boundary as in Fig.~\ref{fig:boundary_details_wrong}(c,d).
In the GR phase, the ground state is a product of local states, and is again unique.

\section{Generalization to qudits and $\mb{Z}_N$ gauge theory \label{sec:ZN_generalization}}

\begin{figure}[!ht]
    \centering
    \includegraphics[width=.42\textwidth]{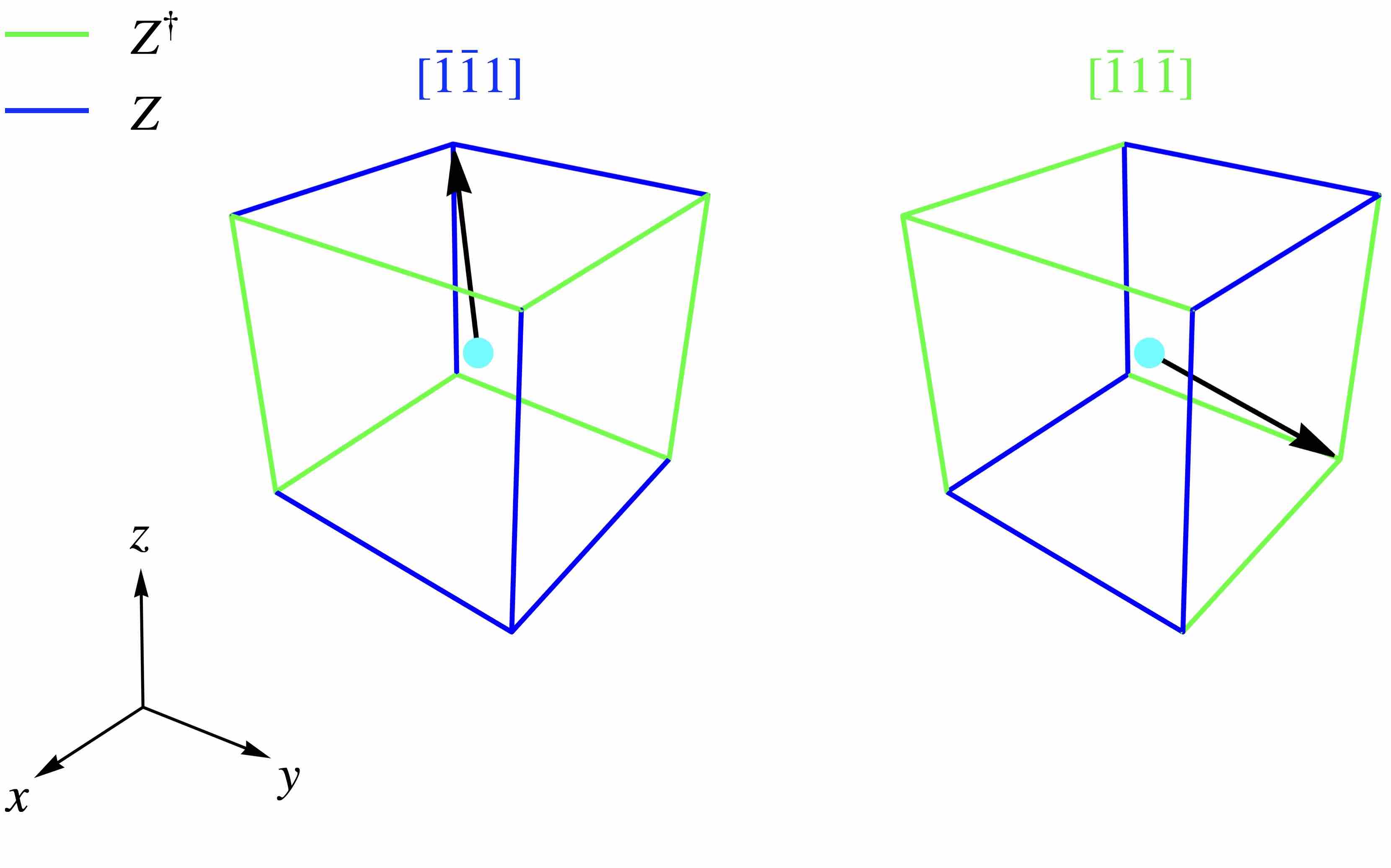}
    \includegraphics[width=.42\textwidth]{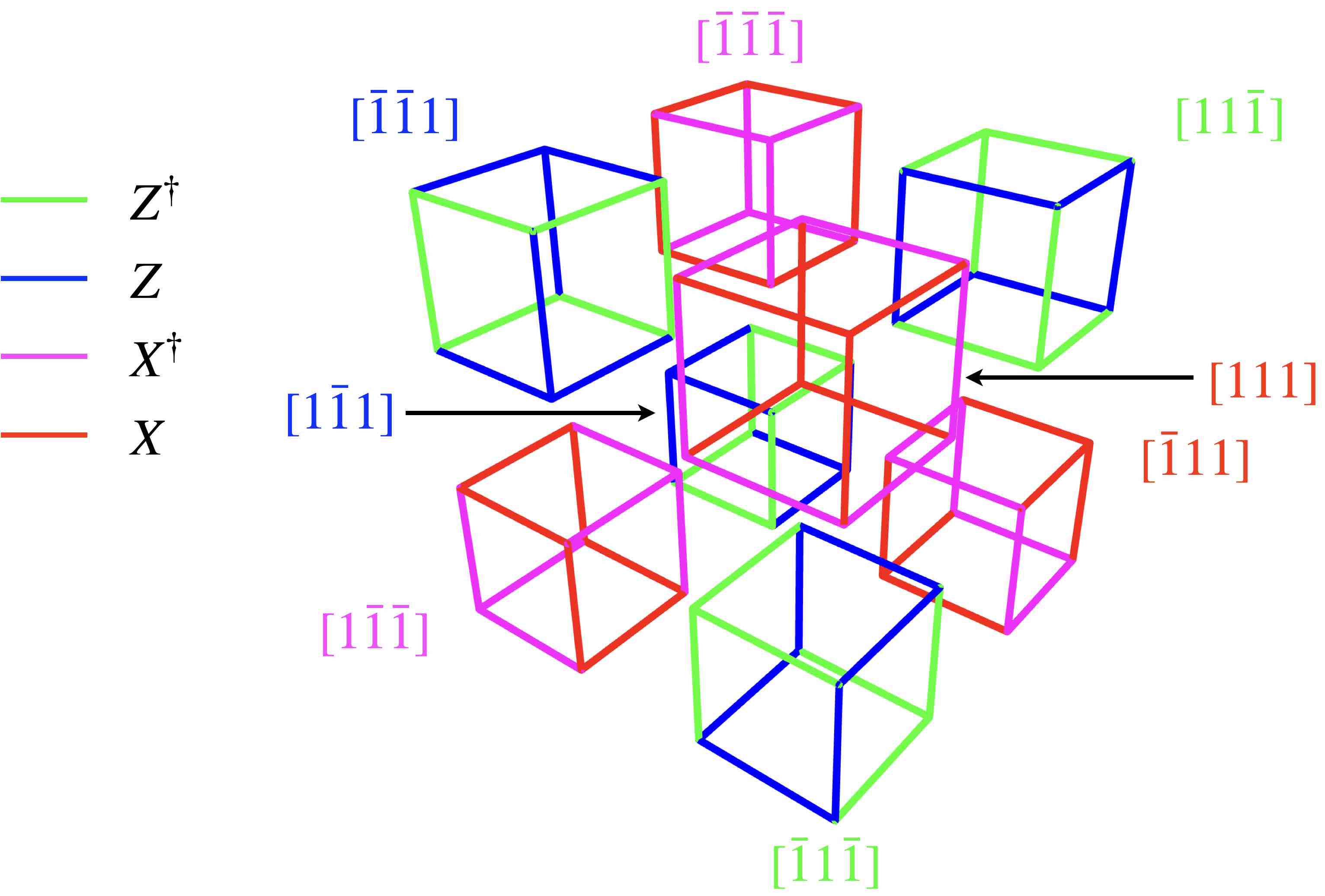}
    \caption{Definition of the vector $u$ for each cell, see the main text and Eq.~\eqref{eq:rule_u_vector_to_cell} for details.
    }
    \label{fig:ZN_code}
\end{figure}

\begin{figure}[!ht]
    \centering
        \includegraphics[width=.45\textwidth]{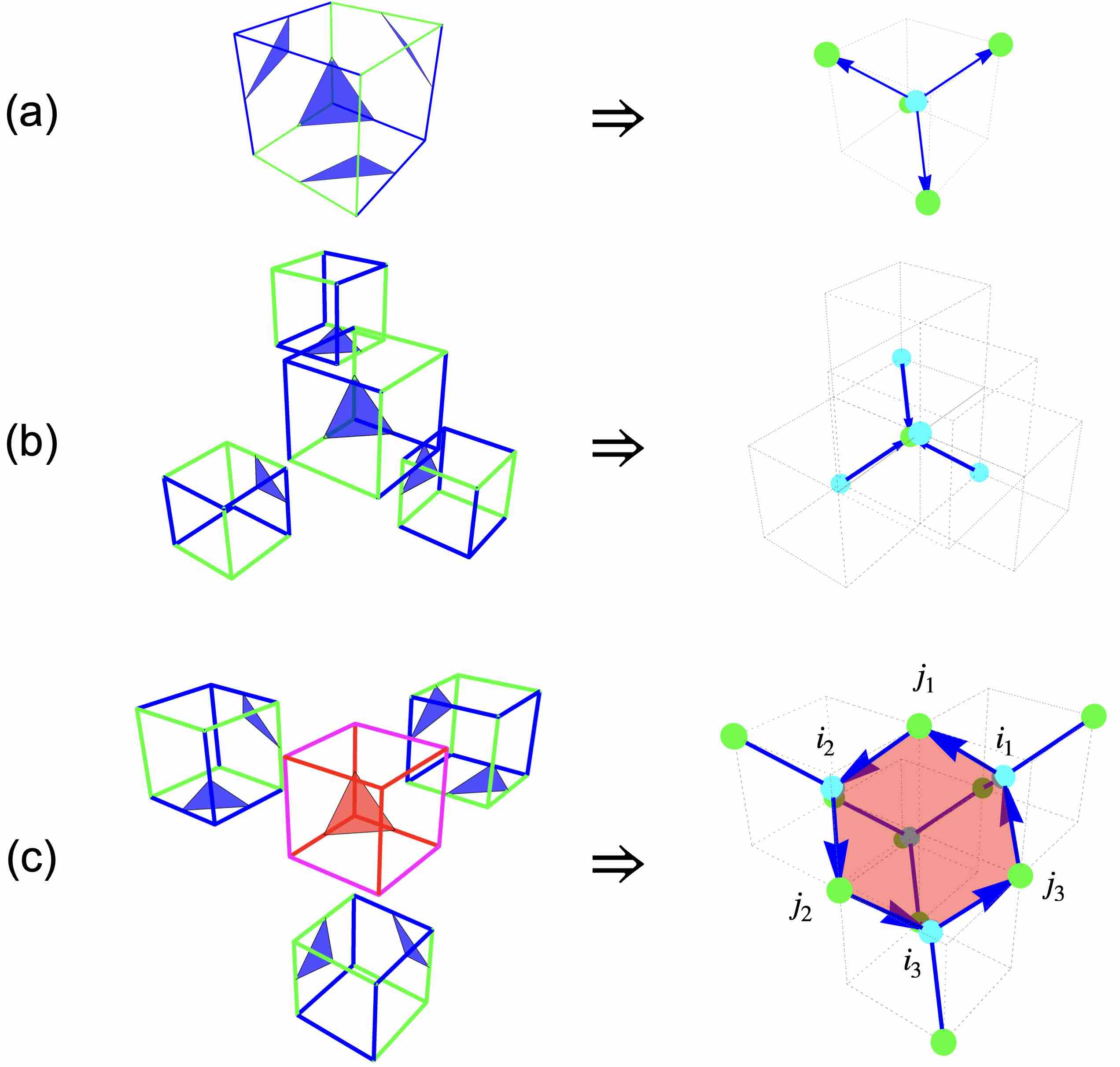}
    \caption{Lattice duality transformation for the Hamiltonian in Eq.~\eqref{eq:Hamiltonian_J_K_ZN}, generalizing Fig.~\ref{fig:face_to_bond}.
    The dualities are described in detail in Eqs.~(\ref{eq:ZN_plaquette_bond_duality}, \ref{eq:ZN_GaussLaw_cyan}, \ref{eq:ZN_GaussLaw_lime}, \ref{eq:ZN_red_check_lattice_curl}).
    }
    \label{fig:ZN_face_to_bond}
\end{figure}

Here we introduce a generalization of the subsystem toric code to qudits with a local dimension $N$, where $N$ is a positive integer.
The Pauli matrices on $N$-dimensional qudits are  generalizations of the qubit case,
\begin{align}
    Z \ket{j} = \omega^{j} \ket{j}, \quad
    X \ket{j} = \ket{(j+1) \, \mathrm{ mod}\, N},
\end{align}
and the commutation relations are
\begin{align}
    X^{a} Z^{b} = \omega^{-ab} Z^{b} X^{a}.
\end{align}

The lattice we consider is the same as Ref.~\cite{KubicaVasmer}, and we put one qudit on each bond of the cubic lattice. We similarly associate $Z$- and $X$- type stabilizers on alternating elementary cells.

Let us first focus on $Z$-type cells.
We label each cell by a unit vector, denoted $u$, which can point from the cell center to one of the cell corners.
We denote $u$ by the usual notation in solid state physics, e.g. $u = [111]$.
Given $u$, we can associate a Pauli operator to each of the 12 edges of the cell in the following way.
We first identify 3 edges that meet at the corner $u$, and 3 edges that meet at the corner $-u$.
We put the same Pauli operator $P$ on these 6 edges, which we choose to be $P = Z$ if $u_z = +1$ and $P = Z^\dg$ if $u_z = -1$.
We then associate $P^\dg$ to the remaining 6 edges.
Two such examples are shown in Fig.~\ref{fig:ZN_code}(a), where we use the blue color to denote a $Z$ operator and the green color to denote a $Z^\dg$ operator.
It is clear that a $Z$ cell is uniquely determined by the vector $u$.
As in Ref.~\cite{KubicaVasmer}, we have a $Z$-type check for each cell corner (there are 8 in total), by taking the product of the 3 Pauli operators meeting at the corner.

We can similarly adopt this convention for the $X$-type cells.

We associate to each elementary cell a vector $u(x,y,z)$ in the following way.
\begin{align}
\label{eq:rule_u_vector_to_cell}
    & u(x,y,z) \nn
    =& 
    \begin{cases}
        [(-1)^{x}, (-1)^{x \oplus z \oplus 1}, (-1)^{x \oplus z} ], \text{ if } x\oplus y \oplus z = 0 (Z \text{ cell})\\
        [(-1)^{x\oplus 1}, (-1)^{x \oplus z}, (-1)^{x \oplus z} ], \text{ if } x\oplus y \oplus z = 1 (X \text{ cell})
    \end{cases}
\end{align}
The lattice is thus periodic, with a periodicity of 2 in each direction.
We show part of the lattice where $(x, y, z) \in \{0,1\} \times \{0,1\} \times \{0,1\}$ in Fig.~\ref{fig:ZN_code}(b).
The entire lattice can be obtained by repeating this $2\times 2 \times 2$ unit cell in all three directions (assuming $L$ is an even number).
It can be checked that all stabilizers commute with each other, and all the checks commute with the stabilizers.
To see this, if suffices to check the unit cell in Fig.~\ref{fig:ZN_code}(b).

One thing to notice about this construction is that on each bond of the lattice, $Z$, $Z^\dg$, $X$, $X^\dg$ each appear once.

We can construct a Hamiltonian in a similar fashion as in Eq.~\eqref{eq:Hamiltonian_J_K}, 
\begin{align}
\label{eq:Hamiltonian_J_K_ZN}
    H =&\ - J_Z \sum_{\bigtriangleup_Z} (\bigtriangleup_Z + \bigtriangleup_Z^\dg) - J_X \sum_{\bigtriangleup_X} (\bigtriangleup_X + \bigtriangleup_X^\dg) \nn 
    &\quad - K_Z \sum_{\mbox{\mancube}_Z} (\mbox{\mancube}_Z + \mbox{\mancube}_Z^\dg) - K_X
    \sum_{\mbox{\mancube}_X} (\mbox{\mancube}_X  + \mbox{\mancube}_X^\dg).
\end{align}
Notice that since now the checks and the stabilizers are not necessarily hermitian, we have to explicitly include their Hermitian conjugates~\cite{bullock2007QuditSurfaceCode}.
Nevertheless, we can still enfoce the (nonhermitian) stabilizer constraints $\mbox{\mancube}_Z = \mbox{\mancube}_Z^\dg = \mbox{\mancube}_X = \mbox{\mancube}_X^\dg = 1$ everywhere, as they commute with everything.
In fact, if $\mbox{\mancube}_Z \ket{\psi} = \ket{\psi}$, we also have $\mbox{\mancube}_Z^\dg \ket{\psi} = \mbox{\mancube}_Z^\dg \mbox{\mancube}_Z \ket{\psi} = \mathbb{1} \ket{\psi}$.

The duality transformation proceeds similarly as in the $\mb{Z}_2$ case, where we have a diamond lattice of bonds connecting cell centers and cell corners, and we similarly associate a dual $\mb{Z}_N$ spin variable on each bond of the diamond lattice, which is perpendicular to the checks (compare with Eq.~\eqref{eq:Z2_plaquette_bond_duality})
\begin{align}
\label{eq:ZN_plaquette_bond_duality}
    \tau^x_{ij} = 
    (\tau^x_{ji})
    ^\dg = \bigtriangleup_Z.
\end{align}
Here, we have used colors to highlight our choice of orientation (see Fig.~\ref{fig:ZN_face_to_bond}(a)): we always choose the bonds to point from a cell center (cyan) to a cell corner (lime).
The analogs of Eqs.~(\ref{eq:Z2_GaussLaw_cyan}, \ref{eq:Z2_GaussLaw_lime}) in this convention read (see Fig.~\ref{fig:ZN_face_to_bond}(a,b))
\begin{align}
\label{eq:ZN_GaussLaw_cyan}
    \forall \text{ cyan } i,&\quad \prod_{j \in n(i)} 
    \tau^x_{ij} = 1, \\
\label{eq:ZN_GaussLaw_lime}
    \forall \text{ lime } j,&\quad \prod_{i \in n(j)} 
    \tau^x_{ij}
    =
    \prod_{i \in n(j)} 
    \tau^x_{ji}
    = 1.
\end{align}
Again we have lattice Gauss laws, when interpreting $\tau^x_{ij}$ as $\mb{Z}_N$ electric fields.

We now examine the commutation relation between the $X$ checks and the $Z$ checks.
As before, for each $X$ checks there are six $Z$ checks that do not commute with it (see Fig.~\ref{fig:ZN_face_to_bond}(c)).
When the six $Z$ checks are arranged cyclically on a ring, we have the following commutation relations,
\begin{subequations}
\begin{align}
    \bigtriangleup_X (\bigtriangleup_Z)_1 =&\
    \omega^{-1} (\bigtriangleup_Z)_1 \bigtriangleup_X, \\
    \bigtriangleup_X (\bigtriangleup_Z)_2 =&\ \omega^{+1} (\bigtriangleup_Z)_2 \bigtriangleup_X, \\
    \bigtriangleup_X (\bigtriangleup_Z)_3 =&\
    \omega^{-1} (\bigtriangleup_Z)_3 \bigtriangleup_X, \\
    \bigtriangleup_X (\bigtriangleup_Z)_4 =&\ \omega^{+1} (\bigtriangleup_Z)_4 \bigtriangleup_X, \\
    \bigtriangleup_X (\bigtriangleup_Z)_5 =&\
    \omega^{-1} (\bigtriangleup_Z)_5 \bigtriangleup_X, \\
    \bigtriangleup_X (\bigtriangleup_Z)_6 =&\ \omega^{+1} (\bigtriangleup_Z)_6\bigtriangleup_X.
\end{align}
\end{subequations}
These are ensured by our construction, in particular the coloring pattern.
Thus, in the dual spin language, we would expect (compare Eq.~\eqref{eq:Z2_red_check_lattice_curl})
\begin{align}
\label{eq:ZN_red_check_lattice_curl}
    \bigtriangleup_X =&\  \tau^z_{i_1 j_1} (\tau^z_{i_2 j_1})^\dg
    \tau^z_{i_2 j_2}
    (\tau^z_{i_3 j_2})^\dg
    \tau^z_{i_3 j_3}
    (\tau^z_{i_1 j_3})^\dg \nn
    =&\ 
    \tau^z_{i_1 j_1} \tau^z_{j_1 i_2}
    \tau^z_{i_2 j_2}
    \tau^z_{j_2 i_3}
    \tau^z_{i_3 j_3}
    \tau^z_{j_3 i_1}.
\end{align}
Interpreting $\tau^z_{ij} \sim e^{i a_{ij}}$ where $a_{ij}$ is a gauge field, $\bigtriangleup_X$ takes the following form
\begin{align}
    \bigtriangleup_X \sim e^{i \int_{\hexagon} d\bs{x} \cdot \bs{a}} \approx e^{i\, (\nabla \times \bs{a})_{\hexagon}}.
\end{align}
We can write $(\nabla \times \bs{a})_{\hexagon} \approx B$, a magnetic flux.
The Gauss laws for the $X$ checks becomes the ``no magnetic monopole'' condition, which is automatically satisfied by the dual spins.

Summarizing, we have
\begin{align}
\label{eq:H_LGT_ZN}
    H^{\rm LGT}_{\mathbb{Z}_N} =& -J_Z \sum_{\avg{ij}} (\tau_{ij}^x + (\tau_{ij}^x)^\dg) \nn
    &\quad - J_X \sum_{\hexagon} 
    (
    \tau^z_{i_1 j_1}
    \tau^z_{j_1 i_2}
    \tau^z_{i_2 j_2}
    \tau^z_{j_2 i_3}
    \tau^z_{i_3 j_3}
    \tau^z_{j_3 i_1} + \mathrm{h.c.}).
\end{align}
This is the familiar Wilson form of a $\mb{Z}_N$ gauge theory Hamiltonian, see e.g. Ref.~\cite{horn1979ZN}.

The phase diagram of this model is shown in Fig.~\ref{fig:ZN_phase_diagram}(b).
For $N<5$, the phase diagram is the same as $N=2$, with two phases separated by a first-order transition.
For $N \ge 5$, the model is similar to a $U(1)$ gauge theory, with an intermediate Coulomb phase with deconfined (gapped) $e$ and $m$ excitations and gapless ``photon'' modes~\cite{tHoof1978permanent, elitzur1979ZN, horn1979ZN, ukawa1980ZN}.

\end{document}